\documentclass[a4paper,11pt]{article}
\pdfoutput=1 

\usepackage{jcappub} 

\usepackage[T1]{fontenc} 

\usepackage{url}
\usepackage{subcaption}
\usepackage{xcolor}
\usepackage{longtable}
\usepackage{lscape}
\usepackage{multirow}
\usepackage{soul} 

\newcommand{\msun}{\mathrm{M}_\odot}
\newcommand{\hu}{\,{\rm km \,s^{-1} \, Mpc^{-1}}} %

\newcommand{\msone}{m_{1,s}}
\newcommand{\mstwo}{m_{2,s}}
\newcommand{\mdone}{m_{1,d}}
\newcommand{\mdtwo}{m_{2,d}}
\newcommand{\ms}{m_{s}}
\newcommand{\md}{m_{d}}
\newcommand{\mmin}{m_\mathrm{min}}
\newcommand{\mmax}{m_\mathrm{max}}

\newcommand{\nexp}{{N_\mathrm{exp}}}

\newcommand{\pdet}{p_\mathrm{det}(\theta)}

\newcommand{\dr}{\mathrm{d}}

\newcommand{\change}[1]{{#1}}
\renewcommand{\st}[1]{{}}
\newcommand{\specialcell}[2][c]{%
  \begin{tabular}[#1]{@{}l@{}}#2\end{tabular}}

\title{Current and future constraints on cosmology and modified gravitational wave friction from binary black holes}

\author[a,1]{K.~Leyde\note{Corresponding author.}}
\author[a,b]{S.~Mastrogiovanni}
\author[a]{D.A.~Steer}
\author[a]{E.~Chassande-Mottin}
\author[c]{C.~Karathanasis}

\affiliation[a]{Universit\'e Paris Cit\'e, CNRS, Astroparticule et Cosmologie, F-75006 Paris, France}
\affiliation[b]{Artemis, Universit\'e C\^ote d'Azur, Observatoire de la C\^ote d’Azur, CNRS, F-06304 Nice, France}
\affiliation[c]{Institut de F\'isica d'Altes Energies (IFAE), Barcelona Institute of Science and Technology, Barcelona, Spain}

\emailAdd{kleyde@apc.in2p3.fr}
\emailAdd{mastrosi@apc.in2p3.fr}
\emailAdd{steer@apc.in2p3.fr}
\emailAdd{ecm@apc.in2p3.fr}
\emailAdd{ckarathanasis@ifae.es}

\abstract{Gravitational wave (GW) standard sirens are well-established probes with which one can measure cosmological parameters, and are complementary to other probes like the cosmic microwave background (CMB) or supernovae standard candles. Here we focus on dark GW sirens, specifically binary black holes (BBHs) for which there is only GW data. Our approach relies on the assumption of a source frame mass model for the BBH distribution, and we consider four models that are representative of the BBH population observed so far. In addition to inferring cosmological and mass model parameters, we use dark sirens to test modified gravity theories. These theories often predict different GW propagation equations on cosmological scales, leading to a different GW luminosity distance which in some cases can be parametrized by variables $\Xi_0$ and $n$. General relativity (GR) corresponds to $\Xi_0= 1$. We perform a joint estimate of the population parameters governing mass, redshift, the variables characterizing the cosmology, and the modified GW luminosity distance. We use data from the third LIGO-Virgo-KAGRA observation run (O3) and find $-$ for the four mass models and for three signal-to-noise ratio (SNR) cuts of 10, 11, 12 $-$ that GR is consistently the preferred model to describe all observed BBH GW signals to date. Furthermore, all modified gravity parameters have posteriors that are compatible with the values predicted by GR at the $90\%$ confidence interval (CI). We show that there are strong correlations between cosmological, astrophysical and modified gravity parameters. If GR is the correct theory of gravity, and assuming narrow priors on the cosmological parameters, we forecast an uncertainty of the modified gravity parameter $\Xi_0$ of $51\%$ with $\sim 90$ detections at O4-like sensitivities, and $\Xi_0$ of $20\%$ with an additional $\sim 400$ detections at O5-like sensitivity. We also consider how these forecasts depend on the current uncertainties of BBHs population distributions.}

\begin{document}
\maketitle
\flushbottom

\section{Introduction}
\label{sec: introduction}

The direct detection of gravitational waves (GWs) has produced a wealth of new results spanning a wide range of scientific areas. GW detections have clearly established the existence of a population of binary black holes (BBH) \cite{LIGOScientific:2016aoc, LIGOScientific:2021djp}, which otherwise eluded observations. Other types of GW sources including binary neutron stars (BNSs) \cite{LIGOScientific:2017vwq} as well as composite systems with a neutron star and a black hole (NSBH) \cite{LIGOScientific:2021qlt} have also been detected.
The most recent GW transient catalog (GWTC-3) \cite{LIGOScientific:2021djp} observed with the LIGO-Virgo detectors \cite{det1-aligo2015,det2-aLIGO:2020wna,det3-Tse:2019wcy,det4-VIRGO:2014yos,det5-Virgo:2019juy} has reported nearly 100 compact binary sources. This has led to statistical studies of the population of these sources in terms of their distances, masses and spins \cite{pop-o1o2-LIGOScientific:2018jsj,LIGOScientific:2020kqk,LIGOScientific:2021psn}, as well as to tests of general relativity (GR) (so far no deviations from GR have been observed) \cite{grtest-gwtc-2-LIGOScientific:2020tif,grtest-gwtc-3-LIGOScientific:2021sio}.

GW observations can also constrain cosmological parameters in several different ways. If an 
electromagnetic (EM) signal is detected from the source, then its redshift $z$ can be obtained from spectrometric measurements.
On combining $z$ with the luminosity distance obtained from gravitational wave data, it is possible to measure the Hubble constant $H_0$, as first proposed in \cite{Schutz:1986gp}. This `bright standard siren' method was applied to the BNS merger GW170817 and its counterpart, resulting in an inferred value of $H_0 = 70.0^{+12.0}_{-8.0} \hu$ (notation for the 68$\%$ confidence level) \cite{LIGOScientific:2017adf}. It is remarkable that a  $20\%$ precision was achieved with this single event, showing that bright standard sirens are a complementary and independent method to measure $H_0$ (see e.g.~\cite{Planck:2018vyg,Riess:2019cxk,Wong:2019kwg} for measurements using other methods). 
The accuracy of \cite{LIGOScientific:2017adf} can be further improved if the distance-inclination degeneracy can be broken, for example with astrophysical information about the gamma ray burst jet and its associated afterglow \cite{Hotokezaka:2018dfi}. 
Using GW170817, it was also possible to measure the fractional difference,  ${\cal O}(10^{-16})$, between the propagation speed of GWs and EM-waves at LIGO-Virgo frequencies, which has led to constraints on different modified gravity theories (for a review, see \cite{Ezquiaga:2018btd} and papers within).
Unfortunately, however, such joint observations of electromagnetic and gravitational waves are predicted to be rare in future science runs \cite{Mastrogiovanni:2020ppa,Mochkovitch:2021prz}.

Alternative approaches to determining $z$ have been proposed that do not require a prompt electromagnetic signal, and can thus be applied to all types of GW sources, including BBH mergers. Following \cite{Schutz:1986gp}, a statistical estimate of the redshift can be obtained by 
identifying nearby galaxies in galaxy catalogs as potential hosts of the GW (compatible with its sky localization) \cite{MacLeod:2007jd,Petiteau:2011we,DelPozzo:2011vcw,Gray:2019ksv}.
This method has been applied to the BBH merger GW170814 \cite{DES:2019ccw} and to the event GW190814 \cite{DES:2020nay}, GWTC-2 \cite{hubble-o2-LIGOScientific:2019zcs} and to GWTC-3, resulting in $H_0 = 68^{+13}_{-7}\hu$ \cite{Virgo:2021bbr}.
The GWTC-2 data in combination with galaxy catalog data has also been used to constrain cosmology and modified gravity theories in \cite{Finke:2021aom}. 
A similar idea is to cross-correlate the GW data with large scale structures from spectroscopic galaxy surveys (such as DESI or SPHEREx)  \cite{Diaz:2021pem,Oguri:2016dgk,Mukherjee:2020hyn,Mukherjee:2020mha}. 
Previous work also used this method to obtain forecasts to test modified gravity theories with third generation detectors \cite{Yang:2021qge,Nishizawa:2019rra}.
The degree of completeness of the electromagnetic survey is a key factor for the level of precision of the final redshift and hence cosmological parameter estimate.

In this paper we follow a different avenue, using solely compact binary source GW data, an approach often referred to as ``dark GW sirens''.  Indeed, the
relation  $\md = (1+z)\ms$ between the source frame mass $m_s$ and the detector frame mass $\md$ inferred from GW observations, together with the assumption that BBHs or BNSs each stem from {a unique and universal mass distribution}, can be used to infer the redshift.
%
%
This approach has been used to fit both cosmological and source mass population parameters using BNSs (see \cite{Taylor:2011fs} for 2nd generation (2G) detectors, \cite{Taylor:2012db} for 3G ones), and using BBHs (see \cite{Farr:2019twy} for 2G detectors, \cite{You:2020wju} for 3G ones, and \cite{Ezquiaga:2020tns} for both).
%
Reference \cite{Farr:2019twy} estimates a potential precision for the Hubble parameter $H(z=0.7)$ of $6 \%$ in an O4 like scenario.\footnote{Currently, the fourth observing run (O4) of the LIGO-Virgo-KAGRA collaboration is planned to start at the end of 2022 \url{https://www.ligo.org/scientists/GWEMalerts.php}. }
%
Note that this uncertainty depends on the assumed astrophysical priors such as the redshift distribution of sources. 

In the literature, some papers have started analyzing the potential of GW dark  sirens to test for deviations from GR on cosmological scales. 
In this context, a standard approximation made in the literature is that only the dynamics of perturbations are modified relative to those of GR, so that the background evolution of the scale factor $a(t)$ is that of $\Lambda$CDM cosmology. Focusing on tensor perturbations, the simplest setup is to consider modifications of gravity in which there are still 2 degrees of freedom propagating at the speed of light (at all energy scales), but whose energy dissipates differently relative to that in GR.  In other words, the GW luminosity distance $d^{\rm GW}_L(z)$ differs from the standard EM luminosity distance, see e.g. \cite{Saltas:2014dha,Nishizawa:2017nef,Belgacem:2017ihm}. In  \cite{Belgacem:2017ihm}, $d^{\rm GW}_L(z)$ was parametrized in terms of two constants $\Xi_0$ and $n$; the corresponding form of $d^{\rm GW}_L(z)$ was shown to be a good fit to certain modified gravity theories, including $f(R)$, non-local gravity, and others. In \cite{2000PhLB..485..208D,Deffayet:2007kf,2021arXiv210908748C}, focusing on models with extra spatial dimensions, $d^{\rm GW}_L(z)$ was shown to depend on the comoving screening scale $R_c$ below which GR should be reproduced.  A third, so called $c_M$-parametrization of $d^{\rm GW}_L(z)$ has been proposed in \cite{
Bellini:2015xja,Alonso:2016suf,Bellini:2015wfa,Bellini:2014fua,2021JCAP...02..043M,2019PhRvD..99h3504L}, and models the effect of a time-dependent Planck mass in terms of dark energy content of the universe with a parameter $c_M$ (which vanishes in GR).
Stability of this model has been studied numerically in \cite{Nishizawa:2017nef}.

In this context,  \cite{2021arXiv210405139M} uses the GW dark siren method to constrain $c_M$: more explicitly, the assumption is that
BBHs follow a \textsc{broken power law} mass distribution; then on fixing
cosmological parameters such as the Hubble constant $H_0$, \cite{2021arXiv210405139M} uses the GWTC-2 catalog  \cite{LIGOScientific:2020ibl} together with upper-limits on the stochastic GW background to constrain $c_M$. Reference \cite{2021arXiv211207650M} also fixes cosmological parameters --- in combination with a \textsc{power law + peak} mass model --- and then uses GWTC-3 to constrain models with extra dimensions and a screening scale $R_c$. 
Reference \cite{2021arXiv210908748C} updated this work after the identification of a missing redshift-dependent factor in $d^{\rm GW}_L(z)$.
Finally, \cite{Mancarella:2021ecn} does not fix cosmological parameters, and uses the dark standard siren method for the GWTC-3 events (with several signal-to-noise ratio (SNR) cuts), assuming a \textsc{broken power law} model only, and probes the parameters $\Xi_0$ and $n$. 
As we will stress below, see also \cite{2021JCAP...02..043M}, cosmological parameters are generally degenerate with modified gravity parameters, and so it is important to keep both free in order to avoid biased results.

In this paper we go significantly further in the analysis of GW dark sirens to test modified gravity and cosmology.  Throughout we leave cosmological parameters free, 
and our aims are two-fold. First we want to probe the sensitivity of this method to the assumptions on the underlying mass distribution.
We thus present a new analysis of the GWTC-3 catalog including, for the first time, a model selection of four different mass distributions and three modified gravity parametrisations.
More precisely we study
the  \textsc{broken power law}, \textsc{Multi peak}, \textsc{power law + peak} and \textsc{truncated power law} mass distributions, as well as the three parametrizations of $d^{\rm GW}_L(z)$ outlined above.\footnote{Where possible, we compare our results to those of \cite{Mancarella:2021ecn}, see sections \ref{sec: real data} and \ref{sec: forecast o4 o5}. Note that we are also using independent codes.}
On calculating the associated Bayes factors, we show that the \textsc{multi peak} mass model is preferred over the other considered mass models, and find no evidence for modified gravity.
We forecast which precision can be obtained for the measurement of the modified gravity parameter $\Xi_0$ with hundreds of GW events with future detector sensitivities
 \cite{det1-aligo2015,det2-aLIGO:2020wna,det3-Tse:2019wcy,det4-VIRGO:2014yos,det5-Virgo:2019juy,det6-KAGRA:2020tym,det7-Aso:2013eba}. We study how this measurement depends on the BBHs merger rate distribution uncertainties from current constraints \cite{LIGOScientific:2020kqk} or on our ignorance of the value of the cosmological background parameters. Considering the credible intervals given by the current BBHs population uncertainties, we find that we will be able to constrain GR deviations with 20\% precision with $\sim500$ BBHs detections. We find that our knowledge of the cosmological background parameters does not strongly affect the precision on the GR deviation parameters. Moreover, we show that with the current distance reach, $\Xi_0$ is almost totally degenerate with the BBH redshift rate evolution, while this degeneracy will be broken with the extended reach enabled by future detector sensitivities.

The structure of the paper is as follows.  
Sec.~\ref{sec: beyond GR parametrization} briefly reviews the different modified gravity luminosity distance parametrizations we consider. In Sec.~\ref{sec: bayesian analysis} we introduce the Bayesian method and describe the population models that define the redshift and source frame mass distributions. In Sec.~\ref{sec: real data} the analysis is applied to GWTC-3 \cite{LIGOScientific:2021djp}, and the results, including the posteriors of the modified gravity parameters are discussed.
Sec.~\ref{sec: forecast o4 o5} describes forecasts with the future detectors and hundreds of GW events. We present our conclusions in Section~\ref{sec: conclusions}.

\section{Modified gravity and parametrization of the GW luminosity distance}
\label{sec: beyond GR parametrization}

Motivated by different problems of the standard cosmological model --- such as the late-time accelerated expansion of the Universe --- many theories have been proposed which modify GR on large scales (see \cite{Clifton:2011jh,Deffayet:2013lga,Joyce:2014kja,Koyama:2015vza,Heisenberg:2018vsk} for reviews). In this paper, we assume that the background evolution is indistinguishable from a flat $\Lambda$CDM, described by the Hubble constant $H_0$, energy fraction in matter today $\Omega_{\rm m}$ and in dark energy today $\Omega_\Lambda$, and consider theories in which there are still two tensor degrees of freedom as in GR. These generally satisfy a modified propagation equation, which in empty space is of the form (see e.g.~\cite{Saltas:2014dha})
\begin{equation}
\label{eq:friction}
    h''_A + 2 \mathcal{H}( 1 - \delta(\eta)) h'_A + k^2 h_A = 0\, ,
\end{equation}
where $\eta$ is conformal time, $(.)'=\partial_\eta(.)$, $k$ is the comoving wavenumber, $A=(+,\times)$ labels the  two independent polarization components of the GWs, and the comoving Hubble parameter $\mathcal{H} = a' /a$.
Note that we have assumed that GWs propagate at the speed of light $c=1$ at all frequencies. In GR, the function $\delta(\eta)$ vanishes; in modified gravity, it is non-zero, and its explicit form will depend on gravity theory under consideration.\footnote{\change{In the specific case of Horndeski theories, $\delta(\eta)$ is one of the four functions of time which fully characterise linear perturbations around a cosmological background \cite{Bellini:2014fua}. The stability of scalar and tensor perturbations imposes certain conditions on these 4 functions see e.g.~\cite{Gao:2011qe,Bellini:2014fua}, which also must be satisfied by any parametrisation of those functions. See \cite{Denissenya:2018mqs,Nunes:2018zot} for further discussions and concrete examples.}}
Eq.~(\ref{eq:friction}) can be solved using the WKB approximation to obtain the GW amplitude $h_A$ e.g.~\cite{Nishizawa:2017nef,Mastrogiovanni:2020gua,Belgacem:2018lbp}. This  scales as the inverse of the GW luminosity distance \footnote{For simplicity, we use $\delta(z)$ as a shorter replacement of $\delta(\eta(z))$.}
\begin{equation}
\label{eq:dGWDS}
d_{L}^{\mathrm{GW}}(z)= d_{L}^{\mathrm{EM}}(z)\,e^{-\int_0^z \mathrm{d}z' \frac{\delta(z')}{1+z'}}\, ,
\end{equation}
where $d_{L}^{\mathrm{EM}}(z)$ is the standard EM luminosity distance which, in a flat $\Lambda$CDM universe in the matter era (the GW events we consider are at $z \lesssim 1$), is given by
\begin{equation}
    d_{L}^{\mathrm{EM}}(z) = (1+z) \int_0^z \frac{\mathrm{d}z'}{H(z')}
\end{equation}
with Hubble parameter
\begin{equation*}
    H(z) = H_0\sqrt{\Omega_{\rm m}(1+z)^3 + \Omega_\Lambda} 
\end{equation*}
and $\Omega_\Lambda = 1-\Omega_{\rm m}$.
In this paper, following others (see \cite{LISACosmologyWorkingGroup:2019mwx} for a review), we focus on three parametrizations of $d_{L}^{\mathrm{GW}}(z)$.

\subsection{\texorpdfstring{$(\Xi_0,n)$}{} parametrization}
\label{sec:Xi_0}

The first parametrization of $d_{L}^{\mathrm{GW}}$ we consider was proposed in \cite{Belgacem:2018lbp} and is given by
\begin{equation}
\label{eq: def Xi parametrization dl}
    d_{L}^{\mathrm{GW}} = d_{L}^{\mathrm{EM}}\left(\Xi_0 + \frac{1-\Xi_0}{(1+z)^n}\right)\,,
\end{equation}
where both $\Xi_0$ and $n$  are assumed positive. For $z\ll 1$, the two luminosity distances coincide, whereas for $z\gg 1$, $d_{L}^{\mathrm{GW}} \rightarrow \Xi_0 d_{L}^{\mathrm{EM}}$. Notice  that for $\Xi_0<1$, the GW luminosity distance is lower, which means that the GW source can be seen to larger (EM) distances than in GR. Thus, one expects to see a higher number of sources.

The parametrization of Eq.~(\ref{eq: def Xi parametrization dl}) has been shown to be a good fit to a number of different modified gravity theories. In the context of scalar-tensor theories, and in particular Horndeski theories \cite{Horndeski:1974wa,Deffayet:2011gz}, when imposing that the speed of GWs is equal to the speed of light, $\delta$ is related to one of the functions on the Horndeski Lagrangian (more exactly, $G_4(\phi)$, see \cite{LISACosmologyWorkingGroup:2019mwx}). Indeed, physically this function $G_4(\phi)$ leads to a time dependent Planck mass, and is the source of the modified friction term in Eq.~(\ref{eq: def cm GW propagation}). The two parameters $\Xi_0$ and $n$ are then related to the change of this Planck mass at low and high redshifts \cite{LISACosmologyWorkingGroup:2019mwx}. Similar comments are true for DHOST theories \cite{Gleyzes:2014dya,Langlois:2015cwa},  scalar-tensor theories beyond Horndeski with second order equations of motion.
The parametrization in Eq.~(\ref{eq: def Xi parametrization dl}) is also a good fit of the GW luminosity distance for a number of other theories, including $f(R)$ gravity and non-local gravity: we refer the reader to Table 1 of \cite{LISACosmologyWorkingGroup:2019mwx} for a summary expressions for ($\Xi_0,n$) for these different theories (they depend on the parameters in the Lagrangian of the modified gravity theory, but potentially also on $\Omega_{\rm m}$).

\subsection{Extra dimensions}
\label{sec:extraD}

Many models of modified gravity have their origins in higher-dimensional spacetimes, for instance DGP models \cite{2000PhLB..485..208D}. 
A characteristic of these theories is the existence of a new length scale, the comoving screening scale $R_c$. Below this scale, the theory must pass the standard tests of general relativity, thus on scales $d \ll R_c$, we expect to recover $d_L^{\rm GW} \sim d_L^{\rm EM}$. On scales $d \gg R_c$, modifications of gravity can become large and all the extra dimensions are probed by the gravitational field whose potential is therefore scales as $\propto d^{-(D-2)/2}$, where $D$ is the total number of spacetime dimensions. As a result, the relationship between the GW and EM luminosity distances can be parametrised by
 \cite{Deffayet:2007kf,2021arXiv210908748C}
\begin{equation}
    d_L^{\rm GW} = d_L^{\rm EM}\left[1+\left(\frac{d_L^{\rm EM}}{(1+z)R_c}\right)^n \right] ^{\frac{D-4}{2n}}\,,
    \label{eq:dpg}
\end{equation}
where $n$ characterises the stiffness of the transition. Note that we define the parameter $n$ differently with respect to \cite{Deffayet:2007kf}.
In the following $D, R_c$ and $n$ will be taken as a free parameters, to be constrained with GW dark siren data.

\subsection{\texorpdfstring{$c_M$}{} parametrization}
\label{sec:cM}

The last parametrisation of $d_L^{\rm GW}$ we consider has been extensively used in the literature, see e.g.~\cite{Bellini:2015xja,Alonso:2016suf,Bellini:2015wfa,Bellini:2014fua,Mastrogiovanni:2020gua}. It assumes that the modified friction term $\delta(z)$ in Eq.~(\ref{eq:friction}) is proportional to the fractional dark energy density $\Omega_\Lambda(z)$, namely 
\begin{equation}
   \delta(z) = -\frac{c_M}{2} \frac{\Omega_{\Lambda}(z)}{\Omega_{\Lambda}}
   = -\frac{c_M}{2} \frac{1}{(1+z)^3 \Omega_{\rm m} + \Omega_\Lambda}
   \,.
   \label{eq:here}
\end{equation}
Gravity is thus modified at late times when dark energy dominates. This parametrization is not a good description of $f(R)$ models \cite{Linder:2016wqw}. However, the advantage of (\ref{eq:here}) is its simple parametrization in terms of a single constant $c_M$. 
It then follows from Eq.~(\ref{eq:dGWDS}), see \cite{2019PhRvD..99h3504L}, that
\begin{equation}
\label{eq: def cm GW propagation}
d_L^{\rm GW} = d_L^{\rm {EM}} {\rm{exp}} \left[\frac{c_M}{2\Omega_{\Lambda}} \ln \frac{1+z}{\left[\Omega_{\rm m}(1+z)^3+\Omega_{\Lambda}\right]^{1/3}} \right]\,.
\end{equation}
In the context of bright GW standard sirens with EM counterparts, this parametrization of the luminosity distance has been investigated in  \cite{2019PhRvD..99h3504L,2021JCAP...02..043M}. We will use it for dark sirens below. 


\section{Analysis framework}
\label{sec: bayesian analysis}

Starting from the observed set of $N_{\rm obs}$ GW detections associated with the data $\{x\}=(x_1,...,x_{ N_{\rm obs}})$, such as the measured component masses and luminosity distance, we wish to infer hyperparameters $\Lambda$ that describe the properties of the source population as a whole.

A \textit{hierarchical Bayesian analysis} scheme can be used to calculate the posterior distribution of $\Lambda$ \cite{Mandel:2018mve,2019PASA...36...10T,Vitale:2020aaz}, namely
\begin{equation}
    p(\Lambda|\{x\},N_{\rm obs}) \propto p(\Lambda)
    \,e^{-N_{\rm exp}(\Lambda)} \left[N_{\rm exp}(\Lambda)\right]^{N_{\rm obs}} \prod_{i=1}^{N_{\rm obs}} \frac{\int p(x_i|\theta,\Lambda)p_{\rm pop}(\theta|\Lambda)\text{d}\theta}{\int p_{\rm det}(\theta,\Lambda)p_{\rm pop}(\theta|\Lambda)\text{d}\theta}\,,
    \label{eq:post}
\end{equation}
where $p(\Lambda)$ is a prior on the population parameters, $\theta$ denotes the set of parameters intrinsic to each GW event, such as component spins, masses, luminosity distance, sky position, polarization angle, inclination, orbital angle at coalescence and the time of coalescence, $N_{\rm exp}(\Lambda)$ is the expected number of GW detections for a given $\Lambda$ and a given observing time $T_{\rm obs}$, while $N_{\rm obs}$ is the number of detected events during an observation time $T_{\rm obs}$. 
The GW likelihood  and probability of detecting a GW event with parameters $\theta$ are denoted by $p(x_i|\theta,\Lambda)$ and $p_{\rm det}(\theta,\Lambda)$, respectively.
These expressions will depend on the sensitivity of the detector network. 
Furthermore, $p_{\rm pop}(\theta|\Lambda)$ represents the \textit{population-modeled} prior.
While the numerator in Eq.~\eqref{eq:post} accounts for the uncertainty on the measurement of the binary properties, the denominator correctly normalizes the posterior and includes \textit{selection effects} \cite{Mandel:2018mve}.

The most probable values of $\Lambda$ correspond to the population parameters that best fit the observed distribution of binaries, both in terms of the intrinsic parameters and of the number of events detectable in a given observation time. 
The population-modeled prior $p_{\rm pop}(\theta|\Lambda)$ is central for the hierarchical Bayesian analysis.
When linking the redshift of the GW events with their luminosity distance (as measured from the data), the distribution of the component source frame masses and redshift is particularly important \cite{Mastrogiovanni:2021wsd}. 

The hyperparameters $\Lambda$ include: a set of cosmological background parameters $H_0$ and the matter energy density $\Omega_{\rm m}$, the parameters related to the GW propagation $\Lambda_\alpha$ (see Sec.~\ref{sec: beyond GR parametrization}) and parameters used to describe the population of BBHs in source masses $\Lambda_m$ and in redshift $\Lambda_z$ (see Sec.~\ref{subsec: z distribution}-\ref{subsec: mass distribution}).
In this work, following \cite{Virgo:2021bbr}, we assume that the source frame mass distributions of BBHs masses are independent from their redshift distribution, namely
\begin{equation}
    p_{\rm pop}(m_{1,s},m_{2,s},z|\Lambda)=p(m_{1,s},m_{2,s}|\Lambda_m)\,p(z|\Lambda_z,H_0,\Omega_{\rm m})\,.
\end{equation}
We use phenomenological models for the source mass distribution $p(m_{1,s},m_{2,s}|\Lambda_m)$ and the \change{(dimensionless)} source spatial distribution $p(z|\Lambda_z,H_0,\Omega_{\rm m})$.
Once a population prior is provided, the number of expected detections can be calculated as
\begin{equation}
    N_{\rm exp}(\Lambda)= R_0 T_{\rm obs} \int p_{\rm det}(m_{1,s},m_{2,s},z,\Lambda) \frac{p(m_{1,s},m_{2,s}|\Lambda_m)f(z|\Lambda_z)}{1+z}\frac{\dr V_c}{\dr z}(H_0,\Omega_{\rm m}) \, \dr z \,\dr m_{1,s} \,\dr m_{2,s}\,,
\end{equation}
where $R_0$ is the merger rate density today in units of $\mathrm{Gpc^{-3}\,\mathrm{yr}^{-1}}$ (it defines the number of events per comoving volume per \change{unit} source frame proper time), 
and $V_c$ is the comoving volume. The dimensionless function $f(z|\Lambda_z)$ describes the evolution of the BBH merger rate with redshift.
It is related to the merger rate of the sources as function of the redshift, as we now discuss.

\subsection{Population-modeled priors: redshift}
\label{subsec: z distribution}
\change{We assume the redshift prior is of the form}
\begin{equation}
    p(z|\Lambda_z,H_0,\Omega_{\rm m}) \propto \frac{f(z|\Lambda_z)}{1+z}\frac{\dr V_c}{\dr z}(H_0,\Omega_{\rm m})\,.
\end{equation}
\change{If $f(z|\Lambda_z)$ is constant, all sources are distributed constantly in comoving volume. }
This describes all merging binaries including the ones not observable due to current sensitivities. The observed population can be obtained by accounting for the sensitivity of the detector network, namely by weighing each source with the probability $\pdet$, \change{see for example \cite{Mandel:2018mve} for the explicit procedure}. 
\change{We model the rate evolution function heuristically as}
\begin{equation}
\label{eq: madau dickinson }
    f(z|\gamma,\kappa,z_p) = \left(1+\frac{1}{(1+z_p)^{\gamma+\kappa}}\right)\frac{(1+z)^\gamma}{1+\left(\frac{1+z}{1+z_p}\right)^{\gamma+\kappa}}\,,
\end{equation}
\change{with three parameters that simply describe} an initially increasing rate with an exponent $\gamma$, followed by a decay with an exponent  $-\kappa$ for redshifts larger than the peak redshift $z_p$. \change{The redshift rate happens to be of the same form as the Madau-Dickinson star formation rate \cite{Madau:2014bja}\footnote{\change{Typical parameter values for the star evolution are $\gamma \sim 1.9,\kappa\sim 3.4,z_p\sim 2.4$ \cite{Callister:2020arv}.}}. However, we want to stress that Eq.\,\eqref{eq: madau dickinson } is not intended to model the star formation rate, nor the binary formation rate, but it is a simple model for the binary merger rate.
The wide priors on the merger rate evolution we consider are given in App.~\ref{app:priors_par} and result in a generic redshift distribution that can significantly differ from the one of the star formation rate.}

\subsection{Population-modeled priors: masses}
\label{subsec: mass distribution}
The mass distributions are based on the four phenomenological models used in \cite{LIGOScientific:2020kqk} to describe the primary mass source frame distribution. In essence these models describe the so called primary (heavier) source frame mass distribution as a power law between a minimum and maximum mass with a slope parameter. 
In order to account for possible accumulation points, certain models add overdensities in the mass spectrum governed by additional parameters, as we will now elaborate.

These models are designed to capture the current state of knowledge about the formation of stellar-mass BH. The \textit{pair instability supernovae} process \cite{ober1983evolution,Bond:1984,1985A&A...149..413G,2002ApJ...567..532H,Umeda:2002sv,Scannapieco:2005zq,Kasen:2011eh, Woosley:2016hmi,Barkat:1967zz,Heger:2001cd,Chatzopoulos:2012fz,Chen:2014aya,Spera:2017fyx,Giacobbo:2017qhh,Belczynski:2016jno} foresees that BHs of masses larger than $40-50\, \msun$ cannot be formed. 
Stars with higher masses lose either part of their mass (which motivates the inclusion of an accumulation point) or are entirely disrupted (which motivates the inclusion of a maximum mass). 

The models of the study represent different attempts to fit the currently available data: the \textsc{Truncated Power Law} model, the \textsc{Power law + peak}: a \textsc{Truncated Power Law} model supplemented by a Gaussian distribution, or by two Gaussian peaks that is referred to as the \textsc{Multi Peak} model, and the \textsc{Broken Power Law} model. 
The secondary mass is described by a \textsc{Truncated Power Law} ranging from the same minimum mass as the primary mass distribution, to a maximum value given by the primary mass. 
The simplest model (\textsc{Truncated Power Law}) has sharp cutoffs, while the other three remaining models have a tapering window applied to the low-end boundary of their distribution.
We provide more details about these mass models in App.~\ref{app:priors_par}. 

Overall, the global model including source population, cosmology and modified gravity aspects includes 11 to 20 parameters that are reviewed in App.~\ref{app:priors_par} and Tables \ref{table: priors} and \ref{table: priors mass}.
The following sections present analyses that are based on a combination of the models presented above: 
Sec.~\ref{sec: real data} compares the four mass models, the three modified gravity models and GR, yielding a total of 16 combinations. 
Concerning the forecast of Sec.~\ref{sec: forecast o4 o5}, the analysis is restricted  to the \textsc{power law + peak} mass model in combination with the $\Xi_0$ modified gravity model.

\section{Application to GWTC-3}
\label{sec: real data}

This section presents the results of the joint parameter estimation of the mass distribution, redshift evolution and modified gravity parameters applied to the events of the GWTC-3 catalog \cite{LIGOScientific:2021djp}. 
A total of 48 runs are performed using the four mass models discussed in Sec.~\ref{subsec: mass distribution} combined with four different models for gravity: the baseline model is referred to as ``GR'' based on $\Lambda$CDM cosmology (no modification of GW propagation), and the three modified gravity models, introduced in Sec.~\ref{sec: beyond GR parametrization} with a flat $\Lambda$CDM background, are referred to as $\Xi_0$, $D$ and $c_M$. 
The robustness of the results is evaluated by reproducing the analysis with three different SNR cuts.

The choices of the prior distributions for the mass models are summarized in App.~\ref{app:priors_par}. The three parameters $\gamma$, $\kappa$ and $z_p$ of Eq.\,\eqref{eq: madau dickinson } used to model the merger rate evolution with redshift  are generated from uniform priors as $\gamma\in\mathcal{U}(0,12)$, $z_p\in\mathcal{U}(0,4)$ and $\kappa\in\mathcal{U}(0,6)$. 
We use a uniform prior for $H_0$ compatible with values from CMB \cite{Planck:2015fie} and standard candle super novae measurements \cite{Riess:2019cxk}, while we fix $\Omega_{\rm m}=0.3065$ \cite{Planck:2015fie}. 
As shown in \cite{2021JCAP...02..043M} $H_0$ is expected to be correlated with the GR modification parameters. If $H_0$ and $\Omega_{\rm m}$ are fixed to incorrect values this could potentially bias the estimation of the GR modification parameters. Thus, if a deviation from GR is concluded from the analysis, the assumed cosmological priors should be questioned. This motivates the additional use of ``agnostic'' or wide priors on the cosmological parameters. For the GW propagation parameters, we use a uniform prior in $\Xi_0 \in [0.3,20]$ and $n \in [1,100]$, for $c_M$ a uniform prior $c_M\in [-10,50]$ and for the number of spacetime dimensions a uniform prior in the interval $D\in[3.8,8]$ is applied.
We intentionally  exclude negative values of $c_M<-10$ since otherwise $d_L^{\rm GW}$ could decrease with redshift (for $z$ large).
We have verified that this prior choice along with $z<20$ used in the analysis produces a strictly increasing $d_L^{\rm GW}$ in redshift.

We will see below that our results favor the baseline GR model and obtain posteriors for the modified gravity parameters that overlap with their ``null'' values in GR. 

\subsection{Description of the data set}

Following the selection adopted in the last measurement of cosmological parameters performed by the LIGO-Virgo-Kagra collaboration (see \cite{Virgo:2021bbr} and associated public data release), we select from the third GW transient catalog (GWTC-3) 42 GWs events consistent with a BBH source type with a network SNR $>11$ and an Inverse False Alarm Rate (IFAR) of $>4 \:{\rm yr}$ (taking the maximum value over the detection pipelines) \cite{LIGOScientific:2021djp}. These selection criteria exclude the asymmetric mass binary GW190814 \cite{LIGOScientific:2020zkf} and the events associated with possible BNSs and NSBHs: GW170817, GW190425, GW200105, GW200115, GW190426 and GW190917 \cite{LIGOScientific:2021psn}. 
The event GW190521 \cite{LIGOScientific:2020iuh} challenges our understanding of black hole formation from massive stars, and in particular of the \textit{pulsational pair instability supernova} (PPISN) theory \cite{Woosley:2016hmi,Spera:2017fyx,Giacobbo:2017qhh,Belczynski:2016jno}.
Despite its high mass, it is compatible with the observed population of BBHs and its inclusion does not change the estimation of the cosmological and modified gravity parameters significantly \cite{Virgo:2021bbr,Mancarella:2021ecn}. 
We therefore include it in the analysis.
Given that some events such as GW200129\_065458 show different mass ratio estimates depending on the waveform approximant, we use posterior distributions combined from the \texttt{IMRPhenom}  \cite{Thompson:2020nei,Pratten:2020ceb} and \texttt{SEOBNR}  \cite{Ossokine:2020kjp,Matas:2020wab} waveform families.

In order to check for possible systematics in the computation of selection biases, we also consider different choices for the SNR cut. 
While keeping the IFAR threshold fixed, we use a lower SNR cut of $10$, leading to 60 selected events and a higher SNR cut of $12$, leading to 35 selected events.

\subsection{Results}

\subsubsection{Model selection}

\begin{figure}
    \centering
    \includegraphics[scale=0.8]{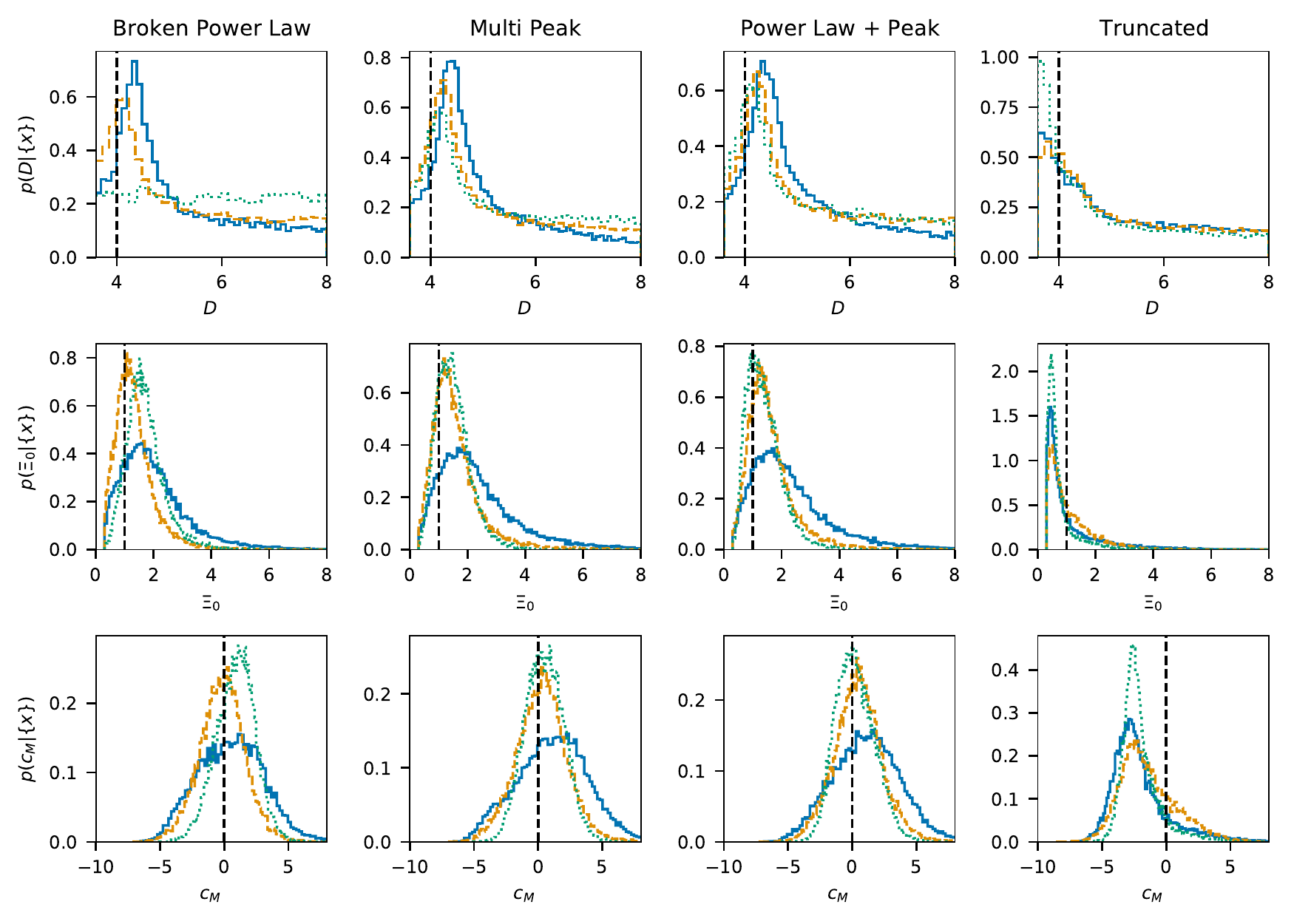}
    \caption{The marginal distributions for the modified gravity parameters $D$ and $\Xi_0$ and $c_M$ for all source mass models and for all three SNR cuts. 
    Blue solid line: result obtained with 42 events with a SNR cut of $11$. Orange dashed line: result obtained with a SNR cut of $12$. Green dotted line: result obtained with an SNR cut of $10$. The vertical black dashed lines indicate the value of the parameter in GR.}
    \label{fig:posteriors_SNR}
\end{figure}

We will begin by discussing our findings in terms of model selection. Table~\ref{tab:bayes} compares the Bayes factors between the different models and the GR+\textsc{Multi Peak} model for a given SNR cut. 
Fig.~\ref{fig:posteriors_SNR} and Tab.~\ref{tab:measure}, give the marginal posteriors of the propagation parameters for the four source mass distributions and the three SNR cuts. 

\begin{table}[t]
  \centering
  \setlength{\tabcolsep}{6pt}
  \renewcommand{\arraystretch}{1.4}
  \begin{tabular}{lcccc}
    \hline
    \\[-5mm]
    \multicolumn{5}{c}{60 BBH events, SNR\;$>10$, IFAR $>4 \:{\rm yr}$} \\[2mm] 
    \hline
    & Broken Power Law& \textbf{Multi Peak}& Power Law + Peak& Truncated\\ 
\textbf{GR}& $-2.4$& $0.0$& $-1.2$& $-6.3$\\ 
$D$& $-2.0$& $-0.2$& $-1.7$& $-6.4$\\ 
$\Xi_0$& $-3.2$& $-0.9$& $-2.1$& $-6.8$\\ 
$c_M$& $-3.0$& $-1.0$& $-2.1$& $-6.5$\\[2mm]
    \hline
    \\[-5mm]
    \multicolumn{5}{c}{42 BBH events, SNR\;$>11$, IFAR $>4 \: {\rm yr}$} \\[2mm] 
    \hline
    & Broken Power Law& \textbf{Multi Peak}& Power Law + Peak& Truncated\\ 
\textbf{GR}& $-1.5$& $0.0$& $-0.8$& $-3.2$\\ 
$D$& $-1.5$& $-0.0$& $-0.9$& $-3.4$\\ 
$\Xi_0$& $-1.9$& $-0.6$& $-1.4$& $-3.9$\\ 
$c_M$& $-1.9$& $-0.9$& $-1.7$& $-3.4$\\[2mm]
    \hline
    \\[-5mm]
    \multicolumn{5}{c}{35 BBH events, SNR\;$>12$, IFAR $>4 \:{\rm yr}$} \\[2mm]
    \hline
    & Broken Power Law& \textbf{Multi Peak}& Power Law + Peak& Truncated\\ 
\textbf{GR}& $-1.2$& $0.0$& $-1.1$& $-2.6$\\ 
$D$& $-1.1$& $-0.4$& $-1.2$& $-2.8$\\ 
$\Xi_0$& $-2.1$& $-1.0$& $-1.9$& $-3.3$\\ 
$c_M$& $-1.9$& $-1.2$& $-1.9$& $-3.1$\\[2mm]
    \hline
  \end{tabular}
  \caption{Logarithm of the Bayes factor normalized to the preferred model (GR + \textsc{Multi Peak}) $\log_{10}\left(\frac{p(\mathrm{data}|\mathrm{gravity}\,\mathrm{model},\,\mathrm{mass}\,\mathrm{model})}{p(\mathrm{data}|\mathrm{GR},\,\mathrm{multi}\,\mathrm{peak})}\right)$ for different SNR cuts assuming a narrow prior for the cosmological parameters (cf.~Table \ref{table: priors} of App.~\ref{app:priors_par}). The preferred BBH mass and modified gravity model are highlighted in bold. The smaller the Bayes factor, the less likely the data can be explained by the given model.}
  \label{tab:bayes}
\end{table}

For all SNR cuts, we find that the preferred model is GR with a \textsc{Multi Peak} source mass model. 
The preference for the \textsc{Multi Peak} model is consistent with \cite{LIGOScientific:2021psn} and \cite{Virgo:2021bbr}. 
This is indicative of the two accumulation points around $10\, \msun$ and $35\, \msun$ in the BBH mass spectrum. 
The high Bayes factor for GR with current GW observations (given the priors on the cosmological parameters) shows that the introduction of additional modified gravity parameters to fit the observed BBH distribution is unnecessary.
This is confirmed by the fact that the marginalized posteriors for the modified gravity parameters (see Fig.~\ref{fig:posteriors_SNR}) are (within the 90\% CI) consistent with their predicted GR values in all cases.
If the observed mass distribution cannot be captured accurately by the mass model, the posteriors of the cosmological and modified gravity parameters can be considerably biased \cite{Mastrogiovanni:2021wsd}.  
We find that at the current sensitivities and the quoted prior on the redshift distribution, the impact of the specific source frame mass model on the $\Xi_0$ measurement is within the 90$\%$ uncertainty. 
Note that the \textsc{Truncated} model is the least preferred model in terms of Bayes factors. This is a consequence of the \textsc{Truncated} model not describing the observed mass spectrum of BBHs. The analysis shows furthermore that the \textsc{Multi Peak} mass model is preferred by a factor of 10  with respect to the simple \textsc{Broken Power Law} model.
See App.~\ref{app: results pl vs plg} for further discussion on the mass model selection and related effect on the estimation of the modified gravity parameters.

\begin{table}[t]
  \centering
  \setlength{\tabcolsep}{6pt}
  \renewcommand{\arraystretch}{1.4}
  \begin{tabular}{lcccc}
    \hline
    \\[-5mm]
    \multicolumn{5}{c}{
    60 BBH events, SNR\;$>10$, IFAR $>4 \:{\rm yr}$} \\[2mm]
    \hline
    & Broken Power Law& Multi Peak& Power Law + Peak& Truncated\\ 
$D$& $5.8^{+2.0}_{-2.0}$& $4.9^{+2.7}_{-1.2}$& $4.8^{+2.8}_{-1.0}$& $4.5^{+3.1}_{-0.8}$\\ 
$\Xi_0$& $1.6^{+1.3}_{-0.8}$& $1.4^{+1.1}_{-0.7}$& $1.3^{+1.2}_{-0.7}$& $0.6^{+1.4}_{-0.2}$\\ 
$c_M$& $1.0^{+2.3}_{-2.6}$& $0.5^{+2.5}_{-2.4}$& $0.1^{+2.7}_{-2.1}$& $-2.4^{+3.3}_{-1.4}$\\[2mm]
    \hline
    \\[-5mm]
    \multicolumn{5}{c}{42 BBH events, SNR\;$>11$, IFAR $>4 \:{\rm yr}$} \\[2mm]
    \hline
    & Broken Power Law& Multi Peak& Power Law + Peak& Truncated\\ 
$D$& $4.7^{+2.9}_{-0.9}$& $4.6^{+2.6}_{-0.8}$& $4.7^{+2.7}_{-0.9}$& $4.8^{+2.8}_{-1.1}$\\ 
$\Xi_0$& $1.8^{+2.6}_{-1.2}$& $2.1^{+3.6}_{-1.4}$& $2.0^{+3.5}_{-1.3}$& $0.7^{+3.0}_{-0.4}$\\ 
$c_M$& $0.5^{+4.1}_{-4.2}$& $1.2^{+4.4}_{-4.8}$& $1.1^{+4.3}_{-4.3}$& $-2.5^{+5.2}_{-2.2}$\\[2mm]
    \hline
    \\[-5mm]
    \multicolumn{5}{c}{35 BBH events, SNR\;$>12$, IFAR $>4 \:{\rm yr}$} \\[2mm]
    \hline
    & Broken Power Law& Multi Peak& Power Law + Peak& Truncated\\ 
$D$& $4.8^{+2.8}_{-1.1}$& $4.6^{+2.9}_{-0.9}$& $4.8^{+2.9}_{-1.0}$& $4.8^{+2.8}_{-1.1}$\\ 
$\Xi_0$& $1.2^{+1.4}_{-0.7}$& $1.4^{+1.8}_{-0.8}$& $1.4^{+1.8}_{-0.8}$& $0.8^{+2.0}_{-0.5}$\\ 
$c_M$& $-0.1^{+2.8}_{-3.0}$& $0.3^{+3.2}_{-3.3}$& $0.4^{+3.2}_{-3.0}$& $-1.8^{+4.6}_{-2.6}$\\[2mm]
    \hline
  \end{tabular}
  \caption{Median and symmetric 90\% confidence intervals of the modified gravity parameters and the selected mass models. The results are shown for different SNR cuts.}
  \label{tab:measure}
\end{table}

With the nominal SNR cut at 11 and a source frame \textsc{multi peak} mass model, the number of spacetime dimensions is constrained to $D=4.6^{+2.6}_{-0.8}$, the phenomenological $\Xi_0=2.1^{+3.6}_{-1.4}$ and the running Planck mass to $c_M=1.2^{+4.4}_{-4.8}$. Our results concerning $\Xi_0$ with the \textsc{Broken Power Law} mass model are also consistent with the results shown in \cite{Mancarella:2021ecn} at the $1\,\sigma$ level, for an SNR cut of 10, 11 and 12. However, our uncertainties are larger, mostly due the larger assumed prior on the Hubble constant. Excluding the prior of the rate of events $R_0$ and the priors on the cosmological parameters (both of which we assume to be more broad), our variable ranges are almost identical to the ones of \cite{Mancarella:2021ecn}.
Furthermore, different selection criteria using the maximum of SNR among the detection pipelines, lead to a slight variation of chosen events for a SNR cut of 11 when compared to \cite{Mancarella:2021ecn}. 
The $c_M$ model in \cite{2021arXiv210405139M} is constrained as $c_M=-3.2^{+3.4}_{-2.0}$ for GWTC-2. 
The analysis presented here obtains a compatible value with a similar or larger uncertainty (depending on the SNR cut) for GWTC-3. 
Finally, \cite{2021arXiv211207650M} finds that the number of extra spacetime dimensions is $D=3.95^{+0.09}_{-0.07}$ with GWTC-3, when fixing the screening scale to 1\,Mpc. 
However, the model of \cite{2021arXiv211207650M} lacks the correction of the redshift factor $(1+z)$ in the GW luminosity distance discussed in \cite{2021arXiv210908748C}.
Nevertheless, our constraints on the number of extra dimensions $D$ are compatible with \cite{2021arXiv211207650M}.

\begin{figure}
    \centering
    \includegraphics[scale=0.4]{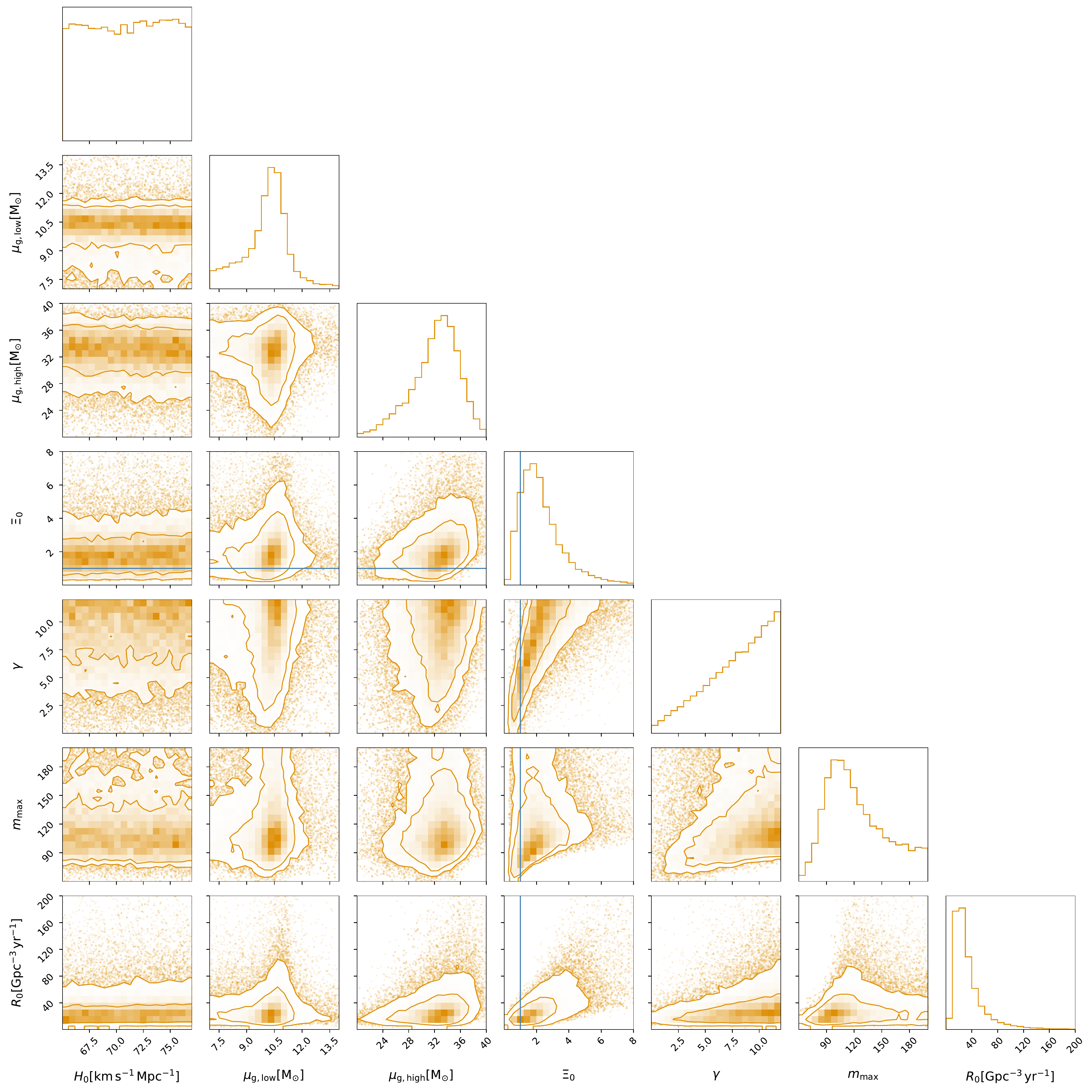}
    \caption{Joint posterior distribution obtained for the \textsc{Multi Peak} mass and $\Xi_0$ models using a GW event catalog produced with a SNR cut at $11$. The blue line indicates the $\Xi_0=1$ value corresponding to GR and the contours indicate the $1\,\sigma$ and $2\,\sigma$ CI.}
    \label{fig:corners_csi}
\end{figure}

\begin{figure}
    \centering
    \includegraphics{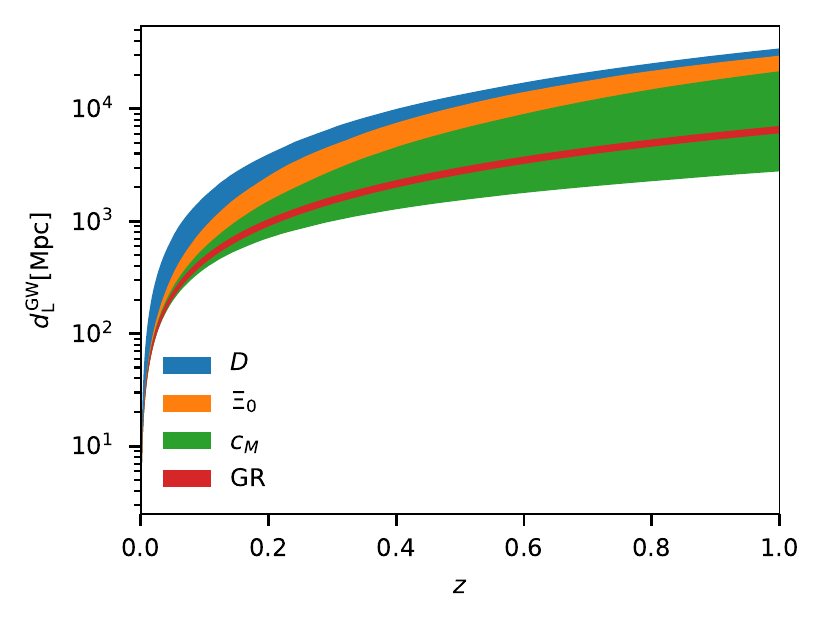}
    \caption{GW luminosity distance vs redshift relation for 42 GW events (SNR cut of 11 analyzed with the \textsc{Multi Peak} mass model). The multicolored bands indicate the 90\% CI.}
    \label{fig:dlconstrains}
\end{figure}

Fig.~\ref{fig:dlconstrains} shows the 90\% CI for the reconstructed GW luminosity distance-redshift relation (using the nominal SNR cut). 
Although the uncertainty levels for the models differ, all predicted GW luminosity distances overlap. Indeed, this is a good validity check of the analysis, since all modified gravity models use identical data. 

One should also note, as shown in Fig.~\ref{fig:dlconstrains}, that conversions of constraints between different modified gravity models are not trivial: When converting variables a prior is implicitly applied; a flat prior on $\Xi_0$ does not correspond to a flat prior on $D$ or $c_M$; one has to include the Jacobian for the change of variable (that will depend on the redshift value at hand).
This complicates the comparison between different propagation models $-$ for instance the rate of events today $R_0$ differs significantly for the $D$ modified gravity model with respect to the $c_M$ and $\Xi_0$ models.
Appendix F of \cite{Mancarella:2021ecn} derives an approximate relationship between the $c_M$ and $\Xi_0$ variables. 

\subsubsection{Correlation between model parameters}
\label{sec:corr_params}
Let us now discuss the parameters of the analysis that most strongly correlate with the modified gravity parameters. 
In Fig.~\ref{fig:corners_csi} we show the joint posterior distribution for the $\Xi_0$ model with the \textsc{Multi Peak} mass model and a SNR cut at $11$. The $\Xi_0$ parameter correlates significantly with the two parameters $\mu_{\rm g, high}$ and $\mu_{\rm g, low}$ which govern the mass features around $10\, \msun$ and $34\, \msun$, and with the rate evolution parameter $\gamma$. These correlations are analogous to the correlations between $H_0$ and $\gamma$, $\mu_{\rm g, high}$ and $\mu_{\rm g, low}$ observed in \cite{Mastrogiovanni:2021wsd,Virgo:2021bbr,Mancarella:2021ecn,2021arXiv211207650M}. 

\paragraph{Degeneracy between $\gamma$ and $\Xi_0$} --- The posterior distribution in Fig.~\ref{fig:corners_csi} shows a strong degeneracy between $\gamma$ and $\Xi_0$. Those two parameters appear to be approximately linearly related. The $\gamma$ vs $\Xi_0$ distribution exhibits a ``ridge'' of about constant height that corresponds to points with a comparable hierarchical likelihood that fit the data equally well. The projection of this ridge onto the $\gamma$ axis shown in the marginal distribution results in the rather high preferred value for $\gamma$.

Consequently, the data appear to be only informative on the ratio $\gamma/\Xi_0$: neither $\gamma$ nor $\Xi_0$ can be robustly measured individually. The constraint on $\Xi_0$ obtained with GWTC-3 is sensitive to the prior set on $\gamma$. As shown with Fig.~\ref{fig:o3 pl vs plg} in App.~\ref{app: results pl vs plg}, larger priors on $\gamma$ correspond to a weaker constraint on $\Xi_0$. However, as detailed in Sec.~\ref{sec: forecast o4 o5}, future detectors will observe much further. This will lead to the breakdown of the $\gamma-\Xi_0$ degeneracy, allowing more robust constraints on the individual parameters (cf.\,Sec.\,\ref{sec:52}).

The same type of degeneracy is observed between $\gamma$ and $c_M$, and thus the same conclusions apply to the marginal distribution obtained with this other gravity model.

\paragraph{Degeneracy between $R_0$ and $\Xi_0$} --- As opposed to \cite{Mastrogiovanni:2021wsd,Virgo:2021bbr} (which are standard cosmological analyses and measure exclusively $H_0$), an extra correlation between $\Xi_0$ and the BBH merger rate density $R_0$ is also observed, as previously noted in \cite{Mancarella:2021ecn}.
The estimation of $R_0$ is related to the expected number of detected events $\nexp$. 
In fact, in the evaluation of the expected number of events, $H_0$ not only modifies the comoving volume as $1/H_0^3$ but also the redshift at which GW events will be detectable (since the SNR depends on the luminosity distance). 
These two effects roughly balance out such that the number of expected detections in a given time is weakly dependent on $H_0$. 
However, this is not the case when considering modified GW propagation. The modified propagation leaves the comoving volume untouched (as it is defined with respect to the EM distance measure) but affects the average redshift at which it is possible to observe GW events. 
As a consequence, the number of expected detections per year strongly depends on the modified gravity parameters.

\begin{figure}
    \centering
    \includegraphics[scale=0.5]{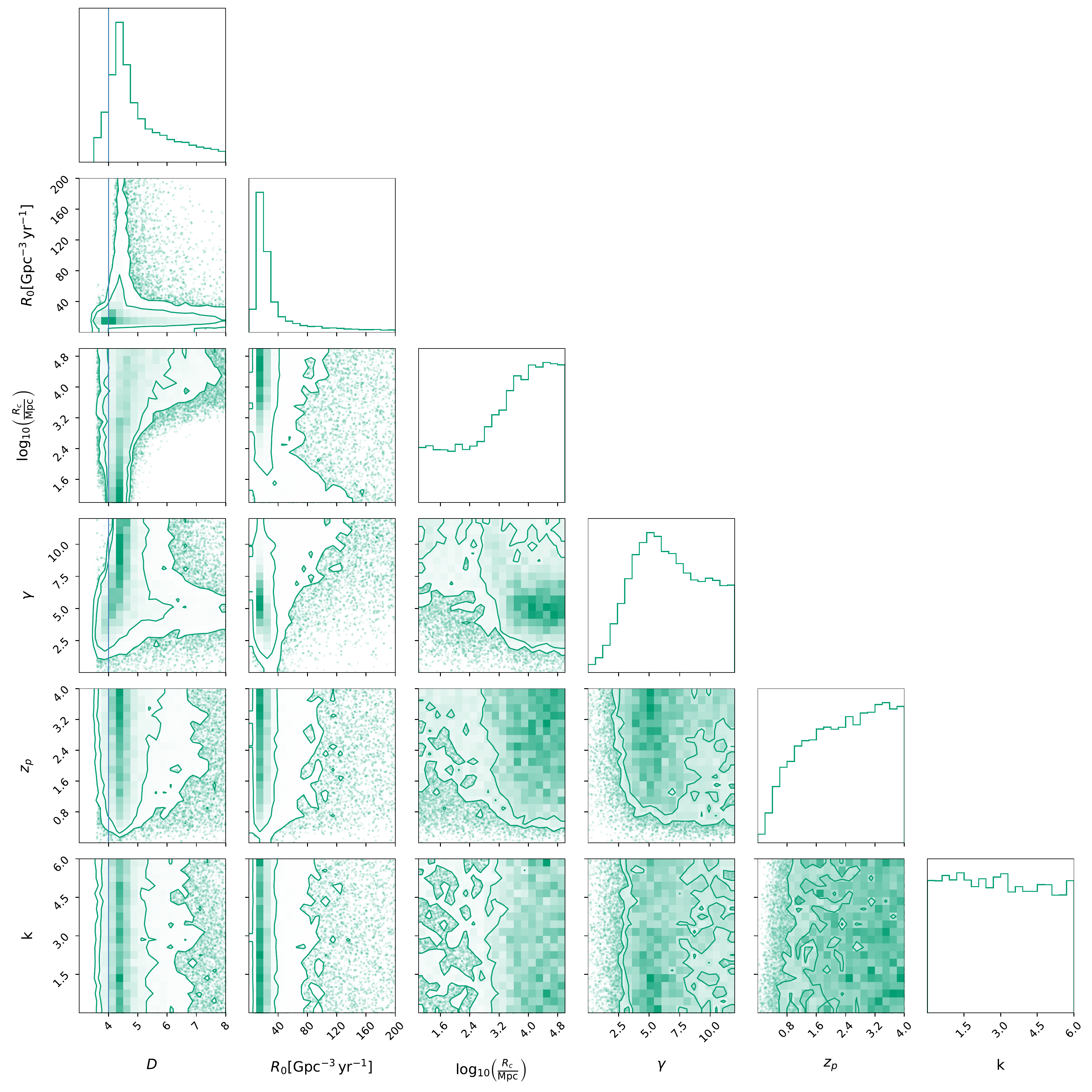}
    \caption{The joint posterior distribution for the \textsc{Multi Peak} mass model and a SNR cut of $11$.
    We show a subset of metaparameters, namely $D$, the GW propagation parameter (extra spacetime dimension), $R_0$, the rate of events today, $R_c$, the comoving screening scale, and $\gamma,z_p$ and $\kappa$, the parameters governing the redshift distribution of sources. }
    \label{fig:corner_D_R0}
\end{figure}

The extent of the correlation between $R_0$ and the modified propagation parameters depends on the assumed propagation model. 
In the case of the $\Xi_0$ model, $R_0$ can be constrained to a value similar to the one provided in \cite{LIGOScientific:2021qlt}. 
This can be compared to Fig.~\ref{fig:corner_D_R0}, where the joint posterior of $R_0$ and the value of spacetime dimensions is shown. 
For this parametrization the correlation between $R_0$ and $D$ is high, affecting the obtained value of $R_0$: The long tail of large $D$ values allows for a tail of high rates $-$ much larger values than in the GR case.  

\section{Forecasts with future GW detections}
\label{sec: forecast o4 o5}
In this section the same analysis is repeated with simulated data sets representative of the upcoming O4 and O5 science runs. For each observation run we assume a $100\%$ duty cycle for each detector. 
The LIGO detectors Hanford and Livingston (HL), are taken at the 'Late High' (O4) and 'Design' (O5) of the aLIGO noise levels \cite{KAGRA:2013rdx}. 
For the Virgo detector (V) we simulate a 'Late High' (O4) and 'Design' (O5) noise curve \cite{KAGRA:2013rdx}.

We draw BBH masses from a fiducial \textsc{Power law + Peak} source frame mass population model (cf.~Eq.~\ref{eq:PLG}), compatible with studies from GWTC-3 \cite{LIGOScientific:2021psn}. 
The parameters of the model are\footnote{The values correspond to the median values obtained from GWTC-2 \cite{LIGOScientific:2020kqk}.} $\alpha=2.63,\beta=1.26,\mmin=4.59\,\msun,\mmax=86.22\,\msun,\mu_g=33.07\,\msun,\sigma_g=5.69\,\msun,\lambda_\mathrm{peak}=0.1,\delta_m=4.82\,\msun$\,. 
This model includes only one sharp feature in the mass spectrum, thus representing a ``pessimistic'' scenario to constrain redshift with mass functions with respect to the \textsc{Multi Peak} model. 
Finally, the redshift evolution is assumed to follow the distribution of Eq.\,\eqref{eq: madau dickinson } with parameters $\gamma=1.9,\,z_p=2.4$ and $\kappa=3.4$, similar to the ones considered in \cite{Callister:2020arv}. 
In particular, the power law exponent for low redshift events $\gamma$ is consistent with the most recent population study of LIGO-Virgo-KAGRA in \cite{LIGOScientific:2021psn}.

Two cases (both with $H_0 = 67.27\hu$, $\Omega_{\rm m}=0.3166$ \cite{Planck:2018vyg}) are analyzed: \textit{(i)} The GR case with no GW propagation modification and \textit{(ii}) A modified gravity model as proposed in \cite{Belgacem:2020pdz}, where $\Xi_0=1.8,n=1.91$. Given the assumed population in redshift and source mass, a BBH merger rate density today of $R_0=23.9\, \mathrm{yr}^{-1}\mathrm{Gpc}^{-3}$ (compatible with current studies) and an observation period of 1 year, the GR case study predicts a total of 87 detections for O4 and 423 detections for O4+O5 with SNR\;$>12$. In the modified gravity scenario, we take the deliberate choice of using the same mass and redshift BBH distributions and fix the same number of the GW detections. This imposes a rescaling of $R_0$ (or equivalently of the observing time) to the larger value\footnote{Though not essential for the present study, it is interesting to note that this value is compatible with the $2\sigma$ CI obtained assuming GR in \cite{LIGOScientific:2021psn}.} of $61.4\, \mathrm{yr}^{-1}\mathrm{Gpc}^{-3}$ since GW signals decay faster when $\Xi_0$ > 1. This choice of fixing the same number of GW data detections has been made to compare the constraints that we would be able to set on $\Xi_0$ in a GR and non-GR case, with the same amount of information from the data.

This section is organized as follows. In Sec.~\ref{sec:51} the technical details of the analysis are provided, in Sec.~\ref{sec:52} we forecast the precision of the modified GW propagation measurement for a GR universe and in Sec.~\ref{sec:53} for a universe with $\Xi_0=1.8$ and $n=1.91$. Finally, Sec.~\ref{sec:54} discusses how current population uncertainties on the BBH distribution impact the forecasts. 

\subsection{Priors and other technical details of the analysis}
\label{sec:51}

For both cases (GR and modified gravity), two analyses are performed: \textit{(a)} A full analysis of joint cosmological, redshift evolution, mass population and modified gravity parameters, using wide priors on the cosmological values. 
\textit{(b)} A joint analysis of all parameters, but fixing the cosmology to measurement uncertainties from other probes such as the CMB \cite{Planck:2018vyg}.  
The priors used are slightly different to the ones applied in Sec.~\ref{sec: real data}, as reported in Tab.~\ref{table: priors} and Tab.~\ref{table: priors mass}.

In order to evaluate the  high-dimensional function $p(\Lambda|\{x\},N_{\rm obs})$ efficiently, the inference library \texttt{bilby} \cite{Ashton:2018jfp} and its Ensemble Monte Carlo Markov Chain implementation are used. The degree of convergence can be verified by studying the auto-correlation times of the sampling chains. 
We use 32 walkers with 50,000 to 100,000 convergence steps and we discard between 1 to 3 times the integrated auto correlation time steps as burn in. 

Another key component is the generation of a proxy for the posterior samples associated with the intrinsic parameters for individual sources in the catalog.
This proxy generation is based on a fit of the typical error made for the detector frame masses and luminosity distance. 
We use a new GW likelihood model calibrated using posteriors obtained from \texttt{bilby}. More details about the new GW likelihood model can be found in App.~\ref{app: likelihood model}.

In the following, all the results are reported at $68\%$ symmetric credible intervals around the median.
Relative uncertainties are computed from the average of the upper and lower sigma interval divided by the median. 

\begin{figure}
    \centering
\begin{subfigure}[t]{.45\linewidth}
  \centering
\includegraphics[width=0.9\linewidth]{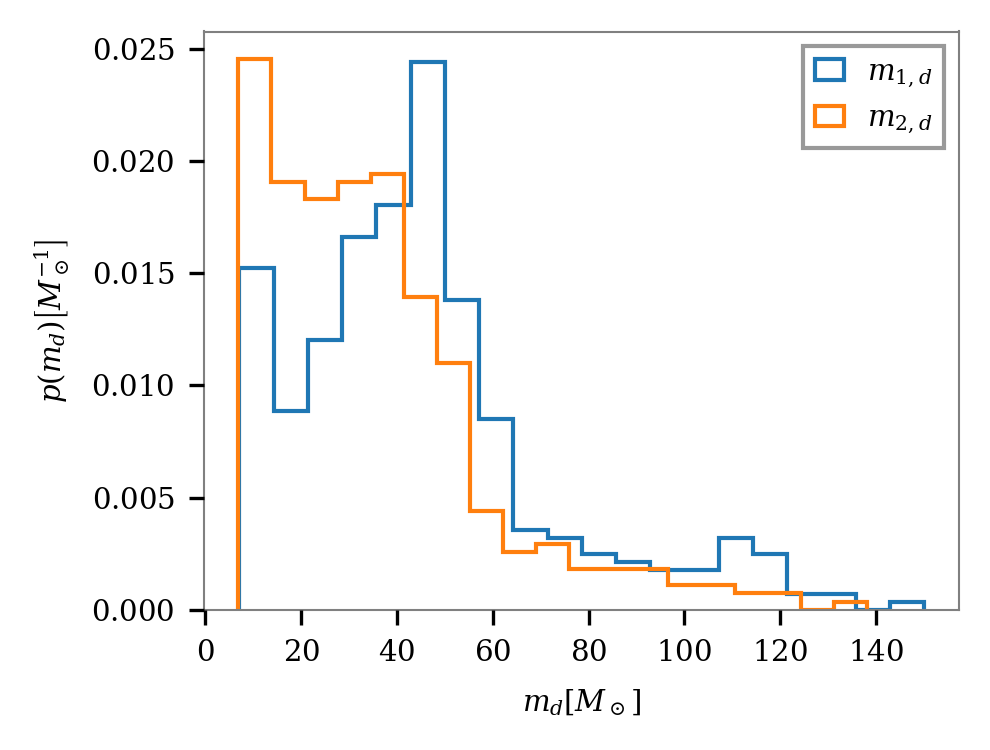}
  \caption{Detector frame mass distribution of the simulated source population. The primary mass is drawn in blue and the secondary mass in orange.  }
  \label{fig:population md o4}
\end{subfigure}%
\hspace{1em}
\begin{subfigure}[t]{.45\linewidth}
  \centering
  \includegraphics[width=0.9\linewidth]{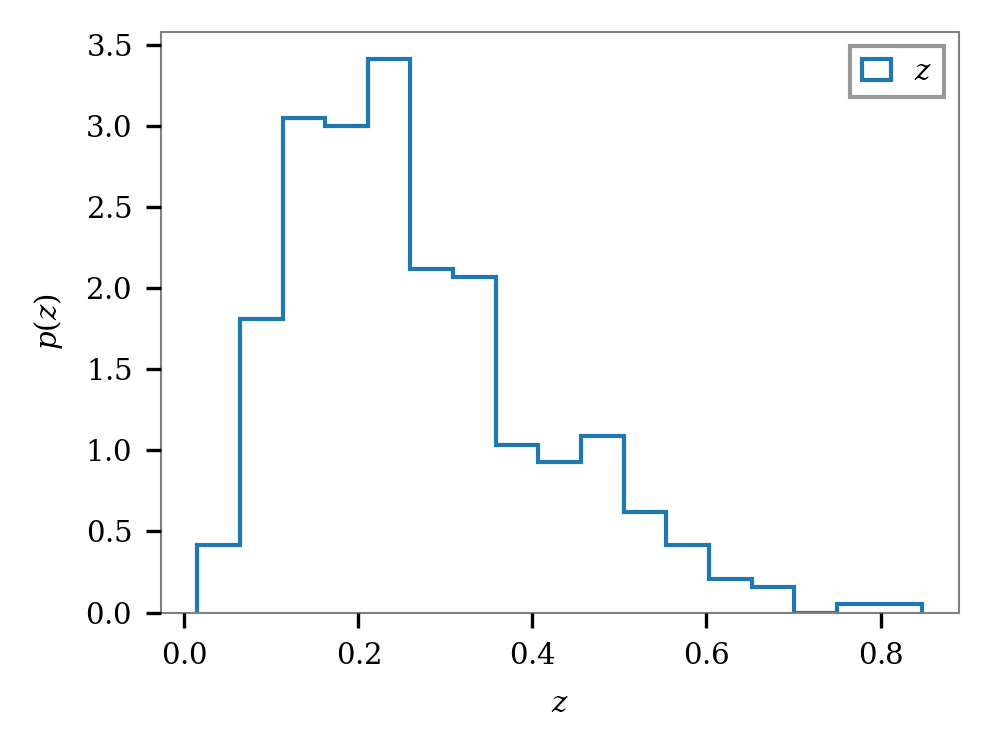}
  \caption{Redshift distribution of the simulated source population.}
  \label{fig:population dL o4}
\end{subfigure}
    \caption{Mass and redshift distributions obtained for a  source population simulated in the case of O4 and assuming GR. The population parameters are $\alpha=2.63,\beta=1.26,\mmin=4.59\,\msun,\mmax=86.22\,\msun,\mu_g=33.07\,\msun,\sigma_g=5.69\,\msun,\lambda_\mathrm{peak}=0.1,\delta_m=4.82\,\msun$\,. This event catalog contains 87 events for the O4 run.}
    \label{fig:population o4}
\end{figure}

\begin{figure}
    \centering
\begin{subfigure}[t]{.45\linewidth}
  \centering
\includegraphics[width=0.9\linewidth]{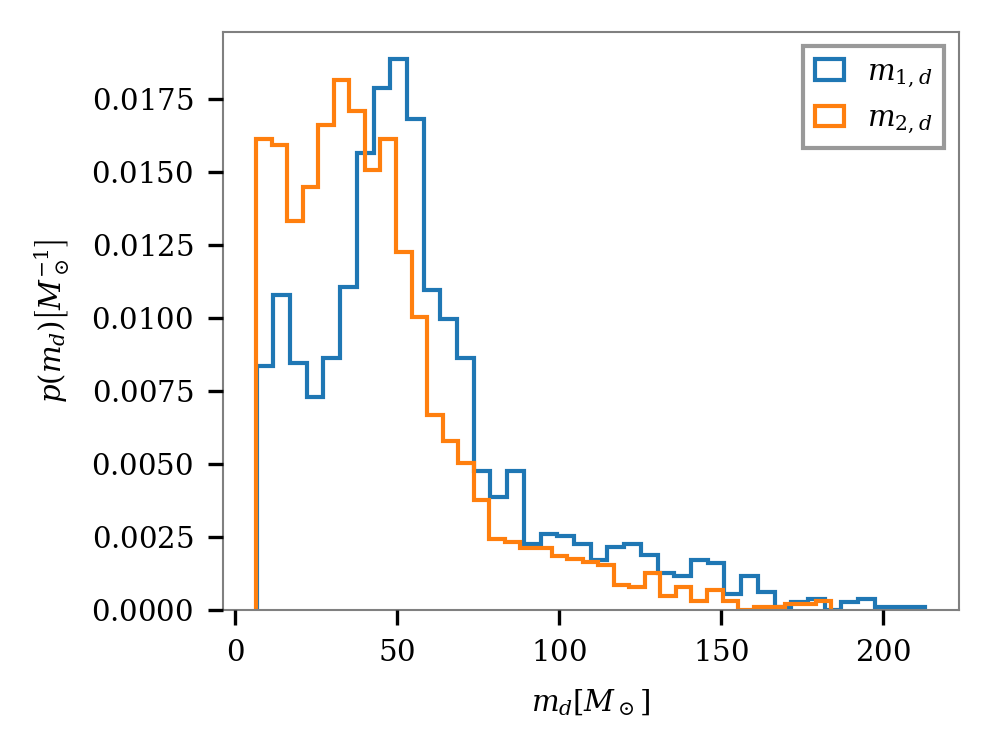}
  \caption{Detector frame mass distribution. The primary mass is drawn in blue and the secondary mass in orange. }
  \label{fig:population md o5}
\end{subfigure}%
\hspace{1em}
\begin{subfigure}[t]{.45\linewidth}
  \centering
  \includegraphics[width=0.9\linewidth]{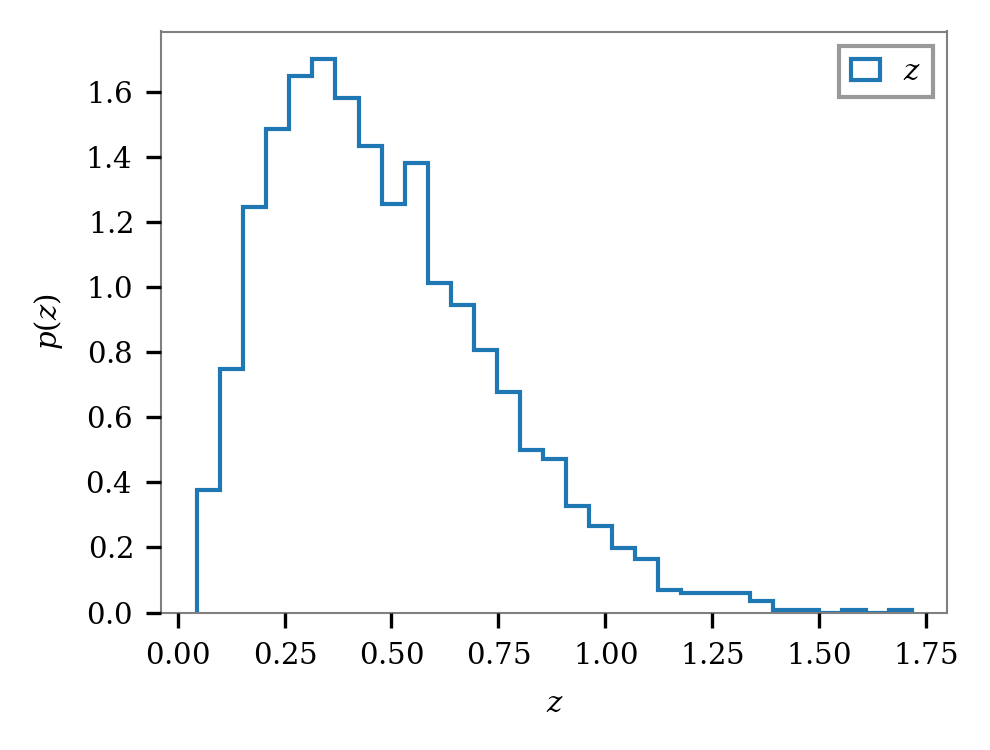}
  \caption{Redshift distribution of the simulated source population. }
  \label{fig:population dL o5}
\end{subfigure}
    \caption{Mass and redshift distributions obtained for a source population simulated in the case of O5 and assuming GR. We use the same parameters of the population as before. The O5 catalog contains 423 events. Compared to the O4 run, the detected events extend much further in redshift.}
    \label{fig:population o5}
\end{figure}

\subsection{Forecasts for a GR Universe}
\label{sec:52} \label{sec:forecast GR case}

The Bayesian population inference scheme described in Sec.~\ref{sec: bayesian analysis} is applied to the simulated data set described above. First, the results of the analysis for wide priors on the cosmological parameters are presented. 

The Hubble constant is constrained at the $42\%$ and $26\%$ precision for O4 and O4+O5, respectively. The matter content $\Omega_{\rm m}$ is essentially unconstrained for all runs. The complete results are shown in Fig.~\ref{fig:o4o5 result GR} of App.~\ref{app: full results agnostic priors} where the posterior distributions obtained with agnostic priors for $H_0$ and $\Omega_{\rm m}$ are shown for all parameters.

Assuming agnostic priors on the cosmological parameters, we obtain $\Xi_0 = 1.40^{+1.03}_{-0.58}$ and $\Xi_0 = 1.27^{+0.41}_{-0.33}$ for O4 and O4$+$O5 combined, respectively. 
The secondary  parameter $n$ that governs modified gravity remains unconstrained, even in the best case with 510 events for O4$+$O5. 
This can be anticipated as, from the construction of the modified gravity model and for the GR case with $\Xi_0=1$, the model is perfectly degenerate under a change of $n$ (cf. Eq.\,\eqref{eq: def Xi parametrization dl}). 
Thus, notwithstanding a very large number of events, $n$ will be unconstrained if $\Xi_0=1$. 

\begin{figure}
    \centering    \includegraphics{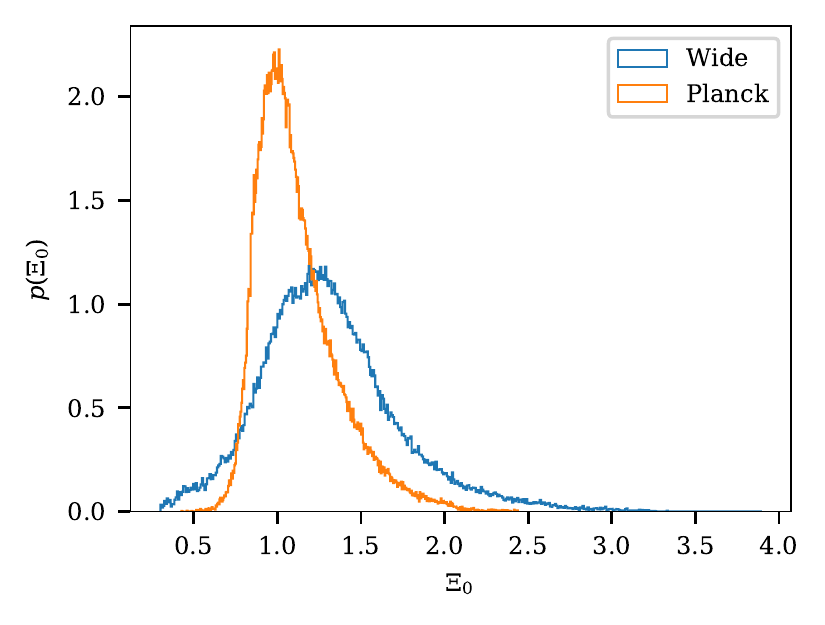}
    \caption{The $\Xi_0$ posterior with 510 events an O4$+$O5 scenario assuming wide priors on the cosmology (blue/\textit{Wide}), or Planck uncertainties (orange/\textit{Planck}). The uncertainty on $\Xi_0$ is reduced by a factor of 1.5 given Planck uncertainties on $H_0$ and $\Omega_{\rm m}$.}
    \label{fig:xi posterior o4o5 comparison}
\end{figure}

Of the redshift evolution parameters $\gamma$, $\kappa$ and $z_p$, the  exponent $\gamma$ for the late redshift evolution is the most constrained, not surprisingly as this parameter relates to the rate evolution at low redshifts. Compared to the previous case using the GWTC-3 events, the joint posterior on $\Xi_0$ and $\gamma$ in Fig.~\ref{fig:o4o5 result GR} appears much more localized around the true value. The event redshift distribution shown in Figure \ref{fig:population dL o5} extends further into redshift than the GWTC-3 data, allowing the $\Xi_0$ vs $\gamma$ degeneracy to be broken.

The $z_p$ posterior is flat for large values and vanishes at lower values of $z$ (the simulated BBH merger rate is strictly incompatible with $z_p=0$ as it increases at low redshifts). 
The early evolution parameter $\kappa$ has a posterior which is almost flat, mostly because few events are detected at redshifts $z>z_p$.
For O4, the maximum mass of the population $\mmax$ has a broad distribution, ranging from $81\,\msun$ to $102\,\msun$ at the 1 sigma confidence level. 
This interval is reduced to $[80,92]\,\msun$ for O4$+$O5 combined. 
Since the mass distributions follow a steeply decreasing power law (with $\alpha=2.63)$, only few sources are informative on the location of the upper mass cutoff $-$ most observed sources have low mass. 

The power law exponent for the redshift evolution $\gamma$ and $R_0$ are strongly degenerate. This is expected as a low rate of events today can be compensated (at first order) by an increase of the number of sources at higher redshifts. 
Furthermore, we find strong correlations between $H_0$ and the characteristic mass scales of the GW population as elaborated previously in \cite{Mastrogiovanni:2021wsd}. 
Since the Hubble constant relates the source's luminosity distance to the its redshift, it shifts the mass distribution to lower or higher values:
A lower Hubble constant places sources generally at lower redshift and thus, source mass and detector frame mass differ less. Conversely, given the measured distribution of detector frame masses, a shift of $H_0$ to larger values can be compensated for by shifting the mass scales to lower values.

We also provide results using restricted priors on $H_0$ and $\Omega_{\rm m}$. As discussed previously, this results in decreased error bars for $\Xi_0$.
Fig.\,\ref{fig:xi posterior o4o5 comparison} compares the marginal posterior of the modified gravity parameter $\Xi_0$ (after O4+O5) applying agnostic priors with applying Planck priors to the cosmological parameters. 
The uncertainty of $\Xi_0$ is reduced by a factor of 1.5 when Planck priors on the cosmological parameters are assumed. 
The uncertainties on the modified gravity parametrization are reduced to $\Xi_0 = 1.47^{+0.92}_{-0.57} $ for O4 and $\Xi_0 = 1.08^{+0.27}_{-0.16} $ for O4$+$O5. Again, $n$ is unconstrained in both observation scenarios.
All the correlations mentioned above apply here as well. 
Even so, since $H_0$ is much better constrained, the correlation $H_0 - \Xi_0$ has a weaker impact $-$ slightly changed cosmological values do not affect the final uncertainties of the variables that parametrize the modified GW luminosity distance.

Compared to \cite{Mancarella:2021ecn}, both in the case of narrow\footnote{Note that we compare the analysis with narrow priors on the cosmological values with the result of \cite{Mancarella:2021ecn} when the cosmological parameters are fixed.} and wide priors on the cosmological values, we obtain $\Xi_0$ constraints after O4$+$O5 twice as worse. 
For wide (narrow) priors, we obtain a relative error of $\sigma_{\Xi_0} = 29\%$ $(20\%)$, whereas the aforementioned source obtains $14\%$ $(10\%)$. 
There are several differences in these analyses.  
With respect to \cite{Mancarella:2021ecn} we assume a two year time span of observation instead of five years, a \textsc{power law + peak} mass model instead of a \textsc{broken power law} model, and a SNR cut of 12 instead of 8. This leads to a large discrepancy in the number of observed events, $\sim4700$ in the aforementioned case, $\sim500$ here. 
Additionally, we apply a uniform in logarithm prior to the rate $R_0$, whereas \cite{Mancarella:2021ecn} uses a uniform prior.
Due to the those different assumptions, it is difficult to trace the origin of this discrepancy. 

When the modified gravity parameters are set to their corresponding GR values ($\Xi_0=1$), the Hubble constant is constrained to $39\%$ and $24\%$ (for the $1\,\sigma$ uncertainty) for O4 and O4$+$O5, respectively. This can be compared to the uncertainty of the Hubble constant of $67\%$ recently reported in \cite{Virgo:2021bbr} from 47 GW dark sirens. The full results with no modified gravity parametrization (and a comparison to previous works) are provided in Appendix \ref{sec:result_O4-O5_GR}.

\subsection{Assessing deviations from general relativity with a \texorpdfstring{$\Xi_0=1.8,n=1.91$}{} universe}
\label{sec:53}

We will now depart from the GR scenario and present the results for a universe in which the GW luminosity distance follows the modified relation of Eq.\,\eqref{eq: def Xi parametrization dl} with $\Xi_0=1.8,n=1.91$.
We simulate a mass and redshift distribution identical to the distribution before (cf.\,the parameters in the caption of Fig.~\ref{fig:population o4}) and we consider again 87 and 423 GW detections as in Sec.~\ref{sec:52}. We choose to retain the same number of events as this allows to see how $\Xi_0$ affects the measurement precision while keeping the same underlying population. Assuming the observing time (or $R_0$) is kept fixed, changing $\Xi_0$ would, in principle, result in a different number of events as $\Xi_0$ changes the observable volume.
Figure \ref{fig:o4o5 result bGR wide} shows the results obtained for O4$+$O5 sensitivities with large priors on the cosmological parameters, leading to $\Xi_0=2.98^{+2.06}_{-1.18}$  for O4 and $\Xi_0=2.31^{+1.07}_{-0.67}$ for O4$+$O5.

For the cosmological parameters in a narrow range consistent with Planck uncertainties in \cite{Planck:2018vyg} the results are as follows: $\Xi_0=2.68^{+2.03}_{-1.01}$ for O4 and $\Xi_0=2.06^{+0.92}_{-0.47}$ for O4$+$O5.
With respect to the previous agnostic priors, these are improvements, of about $6\%$ (O4) and $20\%$ (O4$+$O5) on the full width of the $\Xi_0$ uncertainty. 
After the observation run O5, GR could be excluded at the 2.3 sigma level, if the Universe follows a $\Xi_0=1.8$ model. 

Our precision for $\Xi_0$ from O4$+$O5 corresponds to an increase in uncertainty of $45\%$ and $69\%$, respectively for the case of agnostic and narrow priors on the cosmological parameters, when compared to \cite{Mancarella:2021ecn}. 
While \cite{Mancarella:2021ecn} assumes five years of observation time, the number of observed events is three times as high, since a lower rate of events today is simulated.
Thus, our increased error bars are compatible with the theoretically expected improvement of  $1/\sqrt{3}\approx 58\%$ from the larger number of observed events.
This reasoning is only true when the hierarchical posterior has gaussianized, (see \cite{Mastrogiovanni:2021wsd}). 
Additionally, as we show in Sec.~\ref{sec:53}, the forecast depends on the underlying population and the observed events thereof. 

We end this section by investigating whether these O4$+$O5 constraints can give interesting information on a specific underlying theory. 
In particular we consider the class of quadratic degenerate higher-order scalar tensor (DHOST) theories with action given by \cite{Langlois:2015cwa,BenAchour:2016fzp,Crisostomi:2016czh} 
\begin{equation}
\label{eq: def dhost action}
    S[g_{\mu\nu},\phi] =  \int d^4x \sqrt{-g} \left[ 
F_0(\phi,X) + F_1(\phi,X) \Box \phi + F_2(\phi,X) R + \sum_{I=1}^5 A_I(\phi,X) L_I^{(2)} 
\right] \,,
\end{equation}
where $X=g^{\mu\nu} \partial_\nu \phi\,\partial_\mu \phi $ with the five possible Lagrangians $ L_A^{(2)} $
\begin{align}
    L_1^{(2)} &= \phi^{\mu \nu}\phi_{\mu \nu}\,, \qquad L_2^{(2)} = (\phi_{\nu}{}^{\nu})^2\,, \qquad L_3^{(2)} = \phi_{\nu}{}^{\nu} \phi^\rho \phi_{\rho \sigma} \phi^{\sigma} \,,
\nonumber
\\
 L_4^{(2)} &= \phi^{\mu} \phi_{\mu \nu}  \phi^{\nu \rho } \phi_{\rho}\,, \qquad L_5^{(2)} =( \phi^{\rho} \phi_{\rho \sigma} \phi^{\sigma})^2 \,,
\end{align}
with $\phi_{\mu \nu}=\nabla_\nu \nabla_\mu \phi$, and $\phi_\mu = \nabla_\mu \phi$\,. 
For the functions appearing in the action Eq.~\eqref{eq: def dhost action} we take the same form as in \cite{Crisostomi:2018bsp}
\begin{equation}
    F_0= c_2 X\,, \qquad F_1=0\,,\qquad  F_2=\frac{M_0^2}{2}+c_4 X^2\,,\qquad A_3=-8c_4 - \beta\,.
\end{equation}
For stability, the constant $c_2,\,c_4$ and $\beta$ must satisfy \cite{Crisostomi:2018bsp}
\begin{equation}
    c_2>0\,,\quad c_4>0\,,\quad \beta < -\frac{32}{3}c_4\,.
\end{equation}
We chose $M_0$ to be the Planck mass. 
Given these constants, one can calculate the friction term $\delta(\eta)$ \cite{Crisostomi:2018bsp}, and hence an effective $\Xi_0$ and $n$. 
We have considered the parameter ranges $c_2\in[0,60]$, $\beta\in[-40,0]$ and $c_4\in[0,2]$, and found that $|\Xi_0-1|\lesssim0.3$. 
Therefore, with O4$+$O5 sensitivities, we will not be able to constrain the theory any further. 

\subsection{Impact of underlying BBH population on the \texorpdfstring{$\Xi_0$}{} determination}
\label{sec:54}

The precision with which one can infer GW propagation parameters such as $\Xi_0$ is also determined by the true underlying population of BBH mergers, which is currently uncertain. 

To investigate the dependency of the $\Xi_0$ precision on the true underlying BBH population we further simulate 75 different populations with different minimum mass, position of the Gaussian peak, power law slope etc.\,.
These 75 populations are taken at random from the population samples in \footnote{\url{https://dcc.ligo.org/public/0171/P2000434/003/Population_Samples.tar.gz}} of \cite{LIGOScientific:2020kqk}, and a full analysis, where we marginalize over the total rate of events $R_0$, is performed for each. 
The analysis relies on the same detectors and same sensitivities as in the O4$+$O5 scenario mentioned earlier. The event rate $R_0$ (or the observing time) is tuned to keep the number of detected events constant and equal to 87 for O4 and 423 for O5. This choice allows to single out the impact of the underlying BBH distribution on the measurement precision on $\Xi_0$. The cosmological parameters are fixed to the Planck values as before, and we assume $\Xi_0 = 1$ and $n = 0$.  
For each run, a synthetic population is generated based on a randomly chosen sample for the metaparameters (mass and redshift evolution) taken from the posterior samples in \cite{LIGOScientific:2020kqk}.
To speed up the convergence of the Markov Chain, we neglect the uncertainties of detector frame masses and luminosity distance. 
Therefore, these simulations represent a lower limit on the precision that can be reached for $\Xi_0$.
This provides us with the variability and robustness of the precision obtained for $\Xi_0$ with the current population uncertainties.

\begin{figure}
  \centering
    \includegraphics[width=0.8\textwidth]{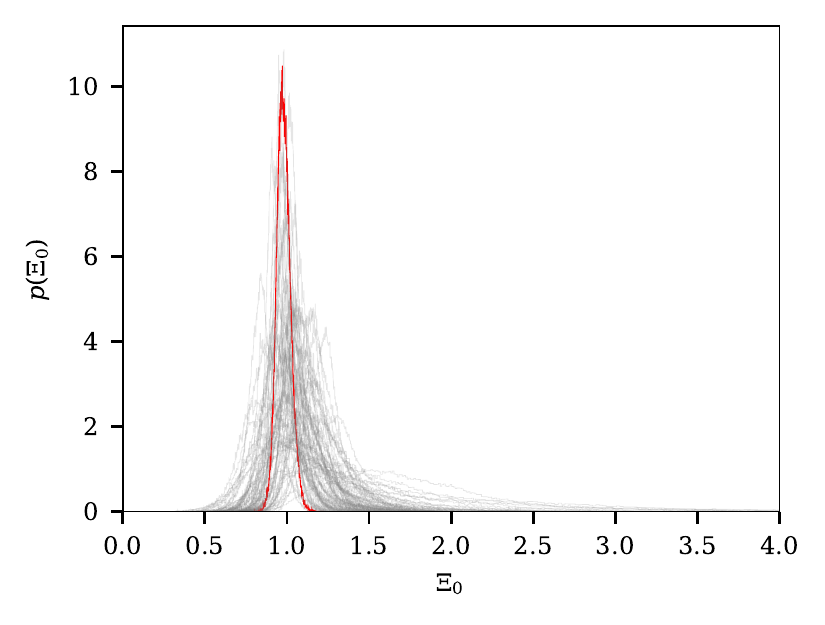}
    \caption{The set of $\Xi_0$ posteriors for various population realizations with different metaparameters, drawn according to the uncertainties in \cite{LIGOScientific:2020kqk}. 
    We have marked the best constrained run (in terms of 1 sigma) in red. 
    We find that half of the runs have an uncertainty of $\sigma_{\Xi_0}\leq12\%$ (1 sigma)). However, of these 75 runs, we find 4 outliers that have significantly larger uncertainties on $\Xi_0$, with $\sigma_{\Xi_0}\geq43\%$. }
    \label{fig:delta run summary}
\end{figure}

Fig.~\ref{fig:delta run summary} presents the distribution of the $\Xi_0$ posterior for the different underlying populations. When averaged over all 75 runs, an uncertainty of $\sigma_{\Xi_0} = 0.16$ is obtained, where $\sigma_{\Xi_0}$ denotes the 1$\,\sigma$ error. 
If we include $90\%$ of the runs (discarding the extreme $5\%$ of all cases), we have a minimum uncertainty of $\sigma_{\Xi_0,\mathrm{min}} = 0.05$ and a maximum uncertainty of $\sigma_{\Xi_0,\mathrm{max}} = 0.43$. We find that $50\%$ of runs, have an uncertainty of $\sigma_{\Xi_0} \leq 0.12 $, while less than 20$\%$ of runs have an uncertainty $\sigma_{\Xi_0} \geq 0.20$.

As shown in App.~\ref{sec:xi0_catalog-realization}, there is a significant variability for different realizations of the same population. The uncertainties given here should not be taken as absolute estimates of the expected errors, since we assume no uncertainties of the intrinsic GW signal parameters.
Indeed, when we compare the average 1$\,\sigma$ interval to the 1$\,\sigma$ interval from Sec.~\ref{sec:forecast GR case} we find a factor of $\mathcal{O}(2)$ difference. 
Thus, the scatter of $1\,\sigma$ confidence levels described above, should be multiplied by a factor of that order.

\section{Conclusions}
\label{sec: conclusions}

This work presented the current and future constraints on cosmology and modified gravity arising from dark GW sirens. We estimated the population distribution of BBHs jointly with cosmological background parameters and parameters governing the modified GW luminosity distance.

The analysis of the most recent publicly released data in \cite{LIGOScientific:2021psn} shows no evidence for deviations from GR at cosmological scales. Current data set constraints on the three theories of gravity considered. Based on an event selection cut at a SNR of 11, in combination of a \textsc{multi peak} source frame mass model, the phenomenological parameter for the GW friction model is constrained to $\Xi_0=2.1^{+3.6}_{-1.4}$, the number of spacetime dimensions to $D=4.6^{+2.6}_{-0.8}$, and the running Planck mass to $c_M=1.2^{+4.4}_{-4.8}$. This study evidences the interdependence of these variables and the parameters that govern the population of BBH sources. 
Compared to existing works, for example \cite{Mancarella:2021ecn}, we have considered a larger number of modified gravity models, and a broader set of mass distributions.
For those, we have explicitly calculated Bayes factors and find no evidence for modified gravity.

The results obtained with the GWTC-3 catalog agree with the current literature reviewed in Sec.\,\ref{sec: real data}, which is indicative of their robustness. 

We also discussed the constraints on GR deviations that one can set with the next science runs O4 and O5. For future observation scenarios we find that the phenomenological parameter $\Xi_0$ can be constrained to $29\%$ (assuming wide priors on the cosmological parameters) and $20\%$ (assuming Planck priors on the cosmology) after O4 and O5 combined.
In the case of a $\Xi_0=1.8$ and $n=1.91$ universe, we find $38\%$ (wide cosmological priors) and $34\%$ (narrow cosmological priors) after O4$+$O5 combined. 
The value of these constraints depend on the simulated population, that is: the true metaparameters, and the specific events observed. To understand this effect, we have generated synthetic populations according to current population knowledge. 
Following O4$+$O5, the $\Xi_0$ posterior has an average uncertainty of $16\%$, if $\Xi_0=1$ and if the intrinsic parameter uncertainties are neglected.

Due to different assumptions on future observation runs (such as different number of observed events, underlying mass models and selection criteria), the forecast of the $\Xi_0$ uncertainties in Sec.~\ref{sec: forecast o4 o5} are difficult to compare to existing results.
However, our results are generically compatible with the result of \cite{Mancarella:2021ecn} in the forecast for modified gravity, see Sec.~\ref{sec:53}. 

Further steps could include the effect of spins, which are expected to be correlated with the mass ratio of the two component masses \cite{Callister:2021fpo}.
It was shown that lensing is likely to become non-negligible for events of O4 and O5 \cite{Ng:2017yiu,Li:2018prc}. 
At the latest for detectors as LISA or third generation observatories such as Einstein Telescope and Cosmic Explorer \cite{Punturo:2010zza,Hall:2019xmm} we expect these effects to become very important due to the longer light paths through the lensing distribution (see \cite{Cusin:2020ezb} for LISA).
Furthermore, one expects the metallicity of stars to change over cosmic history. 
From simulations of the resulting BHs mass distribution \cite{Farmer:2019jed}, this evolution carries over to the position of the pair instability mass accumulation point. 
We have also neglected the time delay of the star formation rate and BBH coalescence.
How these vectors impact the cosmological estimate was recently investigated in \cite{Mukherjee:2021rtw}. We would thus like to extend the present analysis to a mass population that varies with redshift. 
Since we do not use any galaxy catalogs, this analysis may be more robust against calibration uncertainties from galaxy catalogs such as the estimation of its completeness \change{or the estimation of its luminosity function}. 
However, waveform systematics and higher order modes are certain to have an impact of the cosmological parameter uncertainties and we leave it to future work to quantify the expected discrepancies. 
%
%

\clearpage
{\centering\textbf{\huge Appendices}
\vspace{0.1in}}
\appendix

\section{Priors and models}
\label{app:priors_par}
The four phenomenological mass models that we employ (see App.~B of \cite{LIGOScientific:2020kqk}) can be written as a linear combination of two types of statistical distributions. The first is a truncated power law $\mathcal{P}(x|x_{\rm min},x_{\rm max},\alpha)$ described by a slope $\alpha$, and lower and upper bounds $x_{\rm min},x_{\rm max}$ at which there is a hard cutoff

\begin{equation}
\mathcal{P}(x|x_{\rm min},x_{\rm max},\alpha) \propto 
\begin{cases}
    x^\alpha & \left(x_{\rm min}\leqslant x \leqslant x_{\rm max}\right) \\
    0 & \mathrm{Otherwise}.
\end{cases}
\end{equation}
The second is a Gaussian distribution with  mean $\mu$ and standard deviation $\sigma$,

\begin{equation}
\mathcal{G}(x|\mu,\sigma)=\frac{1}{\sigma\sqrt{2\pi}} \exp{\left[ -\frac{(x-\mu)^2}{2\sigma^2}
\right]}\,.
\end{equation}
The source mass priors for the BBH populations that we consider are factorized as 
\begin{equation}
\pi(m_{1,s},m_{2,s}|\Phi_m)=\pi(m_{1,s}|\Phi_m)\pi(m_{2,s}|m_{1,s},\Phi_m),
\end{equation}
where $\pi(m_{1,s}|\Phi_m)$ is the distribution of the primary mass component while $\pi(m_{2,s}|m_{1,s},\Phi_m)$ is the distribution of the secondary mass component given the primary.
For all of the mass models, the secondary mass component $m_{2,s}$ is described with a truncated power law (PL) with slope $\beta$ between a minimum mass $\mmin$ and a maximum mass $m_{1,s}$

\begin{equation}
    \pi(m_{2,s}|m_{1,s},\mmin,\alpha)=\mathcal{P}(m_{2,s}|\mmin,m_{1,s},\beta)\,,
\end{equation}
while the primary mass is described with the several models discussed in the following paragraphs.
For some of the phenomenological models, we also apply a smoothing factor to the lower end of the mass distribution
\begin{equation}
\pi(m_{1,s},m_{2,s}|\Phi_m)=[\pi(m_{1,s}|\Phi_m)\pi(m_{2,s}|m_{1,s},\Phi_m)]S(m_{1,s}|\delta_m,\mmin)S(m_{2,s}|\delta_m,\mmin)\,,
\end{equation}
where $S$ is a sigmoid-like window function that adds a tapering of the lower end of the mass distribution. See Eq.~(B6) and Eq.~(B7) of \cite{LIGOScientific:2020kqk} for the explicit expression of the window function. 

The four phenomenological mass models are summarized in the following list. In Table~\ref{table: priors mass}, we report the prior ranges used for the population hyper-parameters.

\begin{itemize}
    \item \textbf{Truncated Power Law}: The distribution of the primary mass $m_{1,s}$ is described with a truncated power law with slope $-\alpha$ between a minimum mass $\mmin$ and a maximum mass $\mmax$\,. 
\begin{equation}
    \pi(m_{1,s}|\mmin,\mmax,\alpha)=\mathcal{P}(m_{1,s}|\mmin,\mmax,-\alpha)\, .
\end{equation}

    \item \textbf{Power Law + Peak}: The primary mass component is modeled as a superposition of a truncated PL, with slope $-\alpha$ between a minimum mass $\mmin$ and a maximum mass $\mmax$, plus a Gaussian component with mean $\mu_{\rm g}$ and standard deviation
    $\sigma_{\rm g}$,
\begin{align}
    \pi(m_{1,s}|\mmin,\mmax,\alpha,\lambda_{\rm g},\mu_{\rm g},\sigma_{\rm g})=&(1-\lambda_{\rm g})\mathcal{P}(m_{1,s}|m_{\rm min},m_{\rm max},-\alpha)\nonumber\\&+\lambda_{\rm g} \mathcal{G}(m_{1,s}|\mu_{\rm g},\sigma_{\rm g})\,.
    \label{eq:PLG}
\end{align}

    \item  \textbf{Broken Power Law}: The distribution of $m_{1,s}$ follows a PL between a minimum mass $\mmin$ and a maximum mass $\mmax$. The broken power law model is characterized by two PL slopes $\alpha_1$ and $\alpha_2$ and by a breaking point between the two regimes at $m_{\rm break}=\mmin + b (\mmax-\mmin)$, where $b$ is a number $\in [0,1]$. The broken PL model is

    \begin{eqnarray}
        \pi(m_{1,s}|\mmin,\mmax,m_{\rm break},\alpha_1,\alpha_2)&=&\mathcal{P}(m_{1,s}|\mmin,m_{\rm break},-\alpha_1) \nonumber \\ &&
        +\frac{\mathcal{P}(m_{\rm break}|\mmin,m_{\rm break},-\alpha_1)}{\mathcal{P}(m_{\rm break}|m_{\rm break},\mmax,-\alpha_2)}
        \mathcal{P}(m_{1,s}|m_{\rm break},m_{\rm max},-\alpha_2)\,.\nonumber \\ &&
    \end{eqnarray}
    
    \item  \textbf{Multi Peak}: The distribution of $m_{1,s}$ is described as a PL between a minimum mass $\mmin$ and a maximum mass $\mmax$ with two additional Gaussian components with means $\mu_{\rm g,low}, \mu_{\rm g,high}$ and standard deviations $\sigma_{\rm g,low}, \sigma_{\rm g,high}$. The parameter $\lambda_g$ is the fraction of events in the two Gaussian components while the $\lambda_{g,\rm low}$ is the fraction of events in the lower Gaussian component. We denote the set of parameters that describe the \textsc{multi peak} as $\Lambda_{\rm m} = \{\mmin,\mmax,\alpha,\lambda_{\rm g},\mu_{\rm g,high},\sigma_{\rm g,high},\lambda_{\rm g,low},\mu_{\rm g,low},\sigma_{\rm g,low}\}$. 

    \begin{eqnarray}
        \pi(m_{1,s}|\Lambda_{\rm m})&=&(1-\lambda_{\rm g})\mathcal{P}(m_{1,s}|\mmin,m_{\rm break},-\alpha)+
        \nonumber
        \\ &&
        \lambda_{\rm g}\lambda_{\rm g,low} \mathcal{G}(m_{1,s}|\mu_{\rm g,low},\sigma_{\rm g,low})+ \nonumber \\ && \lambda_{\rm g}(1-\lambda_{\rm g,low}) \mathcal{G}(m_{1,s}|\mu_{\rm g,high},\sigma_{\rm g,high})\,.
    \end{eqnarray}

\end{itemize}

\begin{landscape}

\begin{table}
\begin{center}
\renewcommand{\arraystretch}{1.2}
    \begin{tabular}{|c|c|p{3cm} p{3cm}|p{2.8cm} p{2.8cm}|p{4.cm}|} 
\hline
Parameter&Units & \multicolumn{2}{|c|}{Prior GWTC--3 (Sec.~\ref{sec: real data})} & \multicolumn{2}{|c|}{Prior O4$+$O5 (Sec.~\ref{sec: forecast o4 o5})} & Description \\
\hline
 \hline
$\gamma$ &- & \multicolumn{2}{|c|}{$\mathcal{U}(0,12)$} & \multicolumn{2}{|c|}{$\mathcal{U}(0,8)$} & Late redshift evolution in Madau-Dickinson \\
$z_p$ &-& \multicolumn{2}{|c|}{$\mathcal{U}(0,4)$} & \multicolumn{2}{|c|}{$\mathcal{U}(0,6)$} & Characteristic redshift \\ 
$\kappa$ &-& \multicolumn{2}{|c|}{$\mathcal{U}(0,6)$} &\multicolumn{2}{|c|}{ $\mathcal{U}(0,8)$} & Early redshift evolution in Madau-Dickinson\\
\hline
&&&& \textit{GR}& \textit{Modified gravity} & \\
\hline
$R_0$ & $\mathrm{Gpc}^{-3} \mathrm{yr}^{-1}$&\multicolumn{2}{|c|}{$\mathcal{U}(0,1000)$} & Log Uniform$(0.1,100)$ & Log Uniform$(10,300)$ & Event Rate Today\\
\hline \hline
&& \multicolumn{2}{|c|}{\textit{From Planck 2018}~\cite{Planck:2018vyg}} & \textit{Agnostic} & \textit{From Planck} & \\
\hline
$H_0$ &$\hu$ & \multicolumn{2}{|c|}{$\mathcal{U}(65,77)$}& $\mathcal{U}(30,130)$ & $\mathcal{U}(66.07,68.47)$  & Hubble constant \\ 
$\Omega_{\rm m}$  &-& \multicolumn{2}{|c|}{$0.3065$}&  $\mathcal{U}(0.05,0.4)$ & $\mathcal{U}(0.3082,0.3250)$ & Matter content Universe today  \\ 
\hline \hline
$\Xi_0$ &-& \multicolumn{2}{|c|}{$\mathcal{U}(0.3,20)$}& \multicolumn{2}{|c|}{$\mathcal{U}(0.3,10)$} & 
\multirow{2}{*}{\specialcell{Modified gravity\\ parameters, see Eq.\,\eqref{eq: def Xi parametrization dl}}}\\
$n$ &-& \multicolumn{2}{|c|}{$\mathcal{U}(1,5)$}& \multicolumn{2}{|c|}{$\mathcal{U}(1,5)$} &\\
\hline \hline
$D$ &-& \multicolumn{2}{|c|}{$\mathcal{U}(3.8,8)$}& \multicolumn{2}{|c|}{-} & \multirow{3}{*}{\specialcell{Modified gravity \\parameters, see Eq.\,\eqref{eq:dpg}}} \\
$R_c$ &Mpc& \multicolumn{2}{|c|}{Log Uniform$(10,10^5)$}& \multicolumn{2}{|c|}{-} & \\
$n$ &-& \multicolumn{2}{|c|}{Log Uniform$(0.1,100)$}& \multicolumn{2}{|c|}{-} &\\
\hline \hline
$c_M$ &-& \multicolumn{2}{|c|}{$\mathcal{U}(-10,50)$}& \multicolumn{2}{|c|}{-} & \specialcell{Modified gravity\\ parameters, see Eq.\,\eqref{eq: def cm GW propagation}} \\
 \hline
\end{tabular}
 \caption{Overview of the model hyper-parameters along with their associated prior distribution that we assume for the full joint analysis with \texttt{icarogw}. We apply two priors to the cosmological parameters: agnostic priors spanning a wide range of values, as well as informative priors obtained from current Planck measurements. When the O4$+$O5 analysis targets the modified gravity model, the prior distribution on $R_0$ is wider. For the priors of the four different mass models, consider Table \ref{table: priors mass}. }
\label{table: priors}
\end{center}
\end{table}

\end{landscape}

\begin{landscape}

\begin{table}
\begin{center}
\renewcommand{\arraystretch}{1.2}
    \begin{tabular}{|c|c|p{2.3cm} | p{2.3cm}|p{2.3cm} |p{2.3cm} |p{2.3cm}|p{4.4cm}|} 
\hline
Parameter & Units& \multicolumn{4}{|c|}{Prior GWTC--3 (Sec.~\ref{sec: real data})} & \multicolumn{1}{|c|}{Prior O4$+$O5 (Sec.~\ref{sec: forecast o4 o5})} & Description \\
\hline
&& \textsc{Truncated}& \textsc{Broken PL}& \textsc{PL+Peak}& \textsc{Multi Peak}& \textsc{Multi Peak}\\
\hline
 \hline
$\alpha$ &-&$\mathcal{U}(1.5,12)$& n.a.& $\mathcal{U}(1.5,12)$& $\mathcal{U}(1.5,12)$& $\mathcal{U}(1.5,4)$ & (Negative) PL slope of $\mdone$ \\
$\alpha_1$ &-& n.a. & $\mathcal{U}(1.5,12)$& n.a. & n.a. & n.a. & (Negative) PL slope for masses below $m_\mathrm{break}$\\
$\alpha_2$ &-& n.a. &$\mathcal{U}(1.5,12)$ & n.a. & n.a. & n.a. & (Negative) PL slope for masses above $m_\mathrm{break}$\\
$\beta $ &-& $\mathcal{U}(-4,12)$& $\mathcal{U}(-4,12)$& $\mathcal{U}(-4,12)$& $\mathcal{U}(-4,12)$& $\mathcal{U}(0,2)$ &  PL slope of $\mdtwo$ \\ 
$\mmin$ &$\msun$& $\mathcal{U}(2,10)$& $\mathcal{U}(2,10)$& $\mathcal{U}(2,10)$& $\mathcal{U}(2,10)$&$\mathcal{U}(1,10)$ & Minimum mass  \\ 
$\mmax$ &$\msun$& $\mathcal{U}(50,200)$& $\mathcal{U}(50,200)$& $\mathcal{U}(50,200)$& $\mathcal{U}(50,200)$& $\mathcal{U}(60,110)$ & Maximum mass \\ 
$m_\mathrm{break}$ &$\msun$& n.a. & $\mathcal{U}(2,200)$& n.a. & n.a. & n.a. & Mass scale where PL slope changes\\
$\lambda_g$& -&n.a.& n.a.& $\mathcal{U}(0,1)$& $\mathcal{U}(0,1)$ & $\mathcal{U}(0.01,0.5)$ & Fraction of events in Gaussian  \\ 
$\mu_g$&$\msun$ & n.a.& n.a.& $\mathcal{U}(20,50)$& $\mathcal{U}(20,50)$&$\mathcal{U}(10,50)$ & Mean of Gaussian distribution \\
$\sigma_g$&$\msun$ & n.a.& n.a.& $\mathcal{U}(0.4,10)$& $\mathcal{U}(0.4,10)$& $\mathcal{U}(0,40)$ & Standard deviation of Gaussian  \\ 
$\lambda_{\rm g,low}$ &-& n.a.& n.a.&n.a. &$\mathcal{U}(0,1)$& n.a. & Fraction of events in 2$^{\rm nd}$ Gaussian  \\
$\mu_{\rm g,low}$&$\msun$& n.a.& n.a.& n.a.& $\mathcal{U}(7,15)$& n.a. & Mean of the 2$^{\rm nd}$ Gaussian component\\
$\sigma_{\rm g,low}$&$\msun$& n.a.& n.a.& n.a.& $\mathcal{U}(0.4,5)$& n.a. & Standard deviation of the 2$^{\rm nd}$ Gaussian component\\
$\delta_m$&$\msun$ & n.a.& $\mathcal{U}(0,10)$& $\mathcal{U}(0,10)$& $\mathcal{U}(0,10)$&$\mathcal{U}(0,20)$& Smoothing factor at minimum mass \\ 
 \hline
\end{tabular}
 \caption{Priors of the four different mass models for the O3 real data analysis and the forecast of O4$+$O5. The abbreviation n.a. stands for ``non applicable''.  }
\label{table: priors mass}
\end{center}
\end{table}
\end{landscape}

\clearpage
\newpage

\section{Likelihood model}
\label{app: likelihood model}

For a forecast on the estimation of cosmological and the GW luminosity friction parameters under the assumptions of a source mass distribution, we will need to produce extensive GW data. 
In order to reduce computation time, we want to avoid to run a full parameter estimation for the order of a $\sim$90 GW events for O4, and $\sim$400 events for O5. 
Therefore, in this appendix we construct a proxy that mimics the expected measurement uncertainties. 
Since the analysis outlined in Section \ref{sec: bayesian analysis} takes into account the detector frame masses, as well as the GW luminosity distance, we provide samples of these three variables. 

\begin{figure}
    \centering
    \begin{subfigure}[t]{.45\linewidth}
  \centering
\includegraphics[scale=0.9]{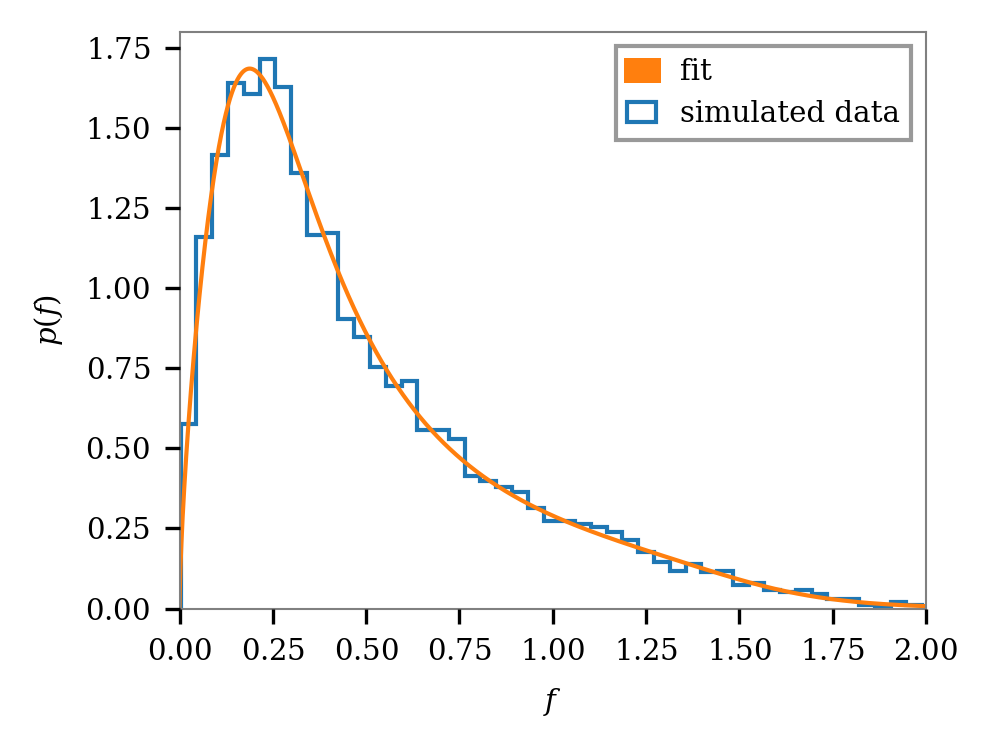}
    \caption{Fit of the geometrical factor $f$ defined in Eq.~(\ref{def: eq f antenna})}
    \label{fig:fit f o4 hlv}
\end{subfigure}%
\hspace{1em}
\begin{subfigure}[t]{.45\linewidth}
  \centering
\includegraphics[scale=0.9]{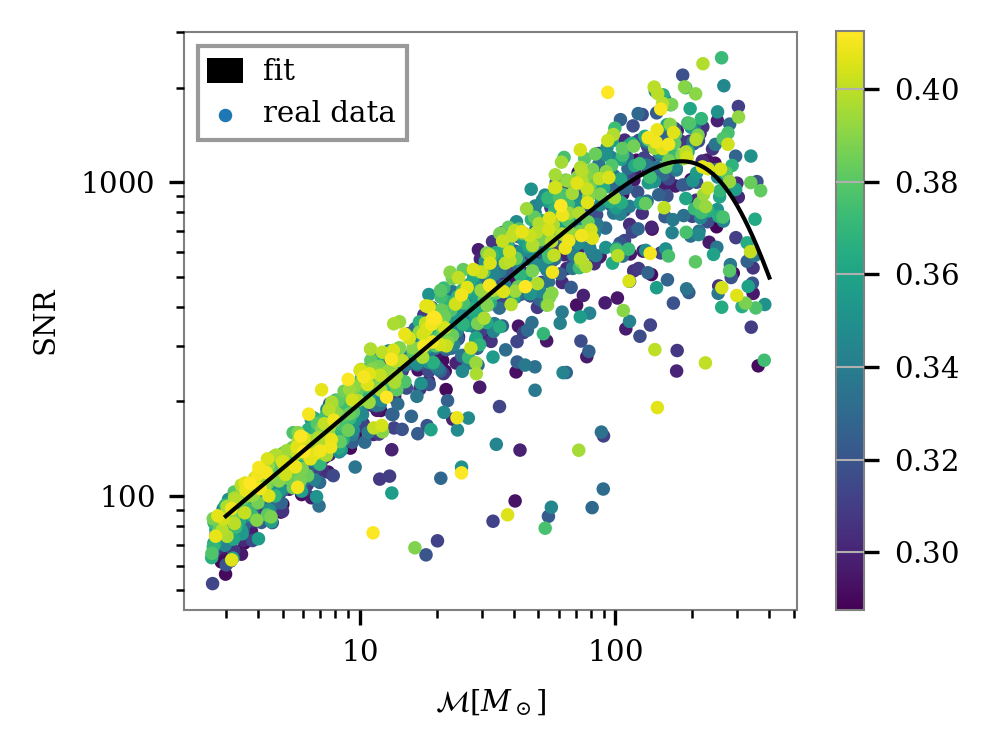}
  \caption{Fit of the SNR in Eq.~\ref{eq: def fit snr} vs chirp mass. The color indicates the geometric factor $f$}
  \label{fig:snr chirp mass o4}
\end{subfigure}%
    \caption{Consistency of the likelihood approximant for the HLV network at O4 sensitivity}
    \label{fig: approximant characterization}
\end{figure}

An approximant should satisfy the following conditions: (i) Capture the uncertainties we expect for a realistic analysis. (ii) Be computationally cheap. 
(iii) Be compatible with the set of detected events used to calculate the selection effect. 
The first point will be verified with the simulation of a full MCMC analysis using \texttt{bilby}, performing a parameter estimation for several GW events. 
The reader will see in the following that point (ii) is fulfilled by construction of the likelihood. We will only consider the amplitude shaping parameters here, namely the component masses, the GW luminosity distance, and $f$, a function of the inclination, sky position and polarization. 
We assume that the SNR $\rho$ takes the following scaling
\begin{equation}
\label{eq: def fit snr}
    \rho= \mathcal{N}\frac{\chi(\mathcal{M})\sqrt{f}}{d_L^{\rm GW}}\,,
\end{equation}
where $\chi$ is a fit, the parameters of which we obtain from simulated data with \texttt{bilby} \cite{Ashton:2018jfp}. This function quantifies the chirp mass dependency of the SNR. We assume a fit of the form 
\begin{equation}
\label{eq: def chi}
\chi(\mathcal{M}) = n\frac{\mathcal{M}^{\alpha_1}}{\mathcal{M}_c^{\alpha_1-\alpha_2} + \mathcal{M}^{\alpha_1-\alpha_2}}\,,
\end{equation}
where the free parameters are $\alpha_1,\alpha_2$ and $\mathcal{M}_c$. The normalization $n$ is redundant of the prefactor $\mathcal{N}$ in Eq.\,\eqref{eq: def fit snr}. This fit interpolates between two PLs above and below a characteristic chirp mass $\mathcal{M}_c$ with exponents of $\alpha_i$.
The fit assumed here is plotted in Figure~\ref{fig:snr chirp mass o4}. 
Finally, the factor $f$ can be seen as a geometrical factor, that takes into account the inclination, polarization angle and sky position of our source
\begin{equation}
    \label{def: eq f antenna}
    f = \sum_{a\in \mathrm{det}} \left[\left(F_+(a) \frac{1 + \cos(\iota)^2}{2}\right)^2 + \left(F_\times(a) \cos(\iota)\right)^2\right]\,,
\end{equation}
where the index $a$ runs over all detectors included in the analysis, $\iota$ is the inclination and $F$ are the antenna response functions.
Again, we fit the $f$ distribution from a set of simulations. The fit of the $f$ distribution can be found in Figure \ref{fig:fit f o4 hlv} for the HLV network at O4 sensitivity. 
Finally, the normalization factor $\mathcal{N}$ is constrained from calculating the number of detected sources with \texttt{bilby}.

For the likelihood model or the distribution of measured values we make the following assumptions:

\begin{itemize}
    \item The square of the measured network SNR follows a non-central chi square distribution with the number of degrees of freedom corresponding to the number of detectors included in the analysis. 
    \item We assume Gaussian distributions on the chirp mass $\mathcal{M}$ and the symmetric mass ratio $\eta$, where the standard deviation is inversely proportional to the network SNR.
    \item For the likelihood distribution of the amplitude factor $f$, we fitted the $f$ distribution for a full (aligned spins) parameter analysis. As a first order approximation, we take this likelihood to be independent of SNR. It is only dependent on the measured value of $f$. 
\end{itemize}

The procedure we take is the following. From our mass and redshift distribution, we draw their corresponding values. From the $f$ distribution as plotted in Figure \ref{fig:fit f o4 hlv} we draw an amplitude. 
This produces a source with (true) parameters $\msone,\mstwo,z$ and $f$. 
Given this event, we can use Eq.\,\eqref{eq: def fit snr} to compute its associated network SNR.
We then draw the \textit{observed network SNR} from a non-central chi square distribution. 
If this SNR passes a threshold of 12, we produce measured values and posterior samples in $f, \mathcal{M},\eta$ and $\rho$ according to our assumptions above.
Inverting chirp mass and symmetric mass ratio posterior samples in $\mdone$ and $\mdtwo$ is straightforward. 
Now that we are given Eq.\,\eqref{eq: def fit snr}, we can also directly invert the posterior samples to obtain samples of the luminosity distance. 
Note that we apply the standard priors that are used in PE, namely a volumetric prior $\left(d_L^{\mathrm{GW}}\right)^2$ and flat priors in the detector frame masses. This implies that the priors in chirp mass and symmetric mass ratio are non-uniform.

\begin{figure}
    \centering
    \begin{subfigure}[t]{.48\linewidth}
  \centering
\includegraphics[scale=0.95]{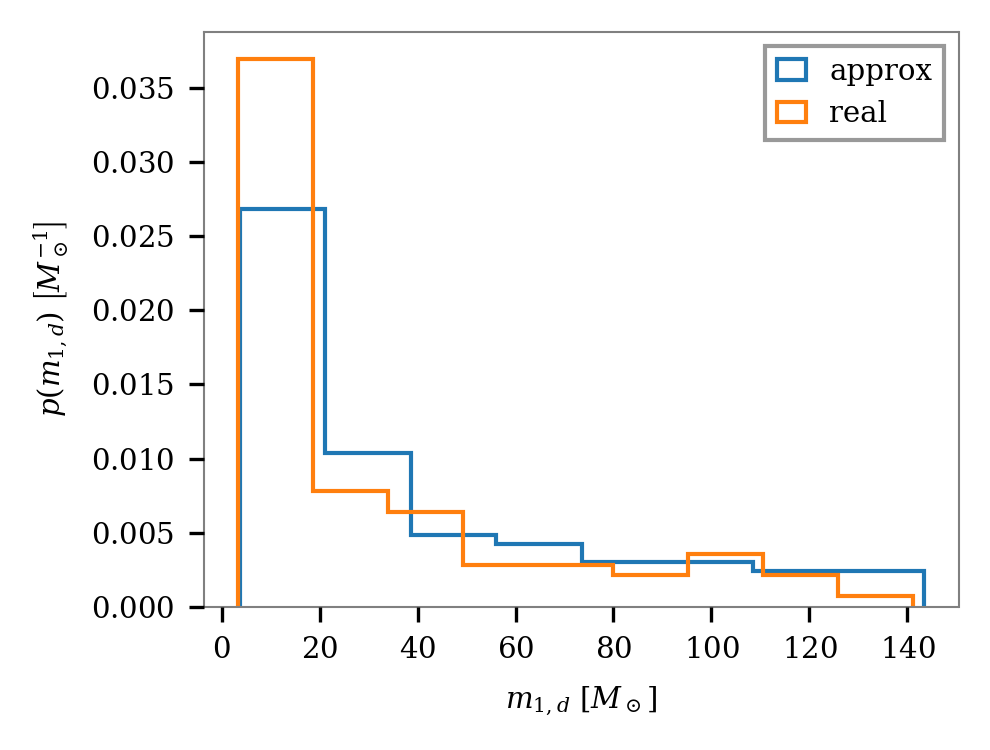}
     \caption{Heavier detector frame mass}
    \label{fig:population dl}
\end{subfigure}%
\hspace{1em}
\begin{subfigure}[t]{.48\linewidth}
  \centering
\includegraphics[scale=0.95]{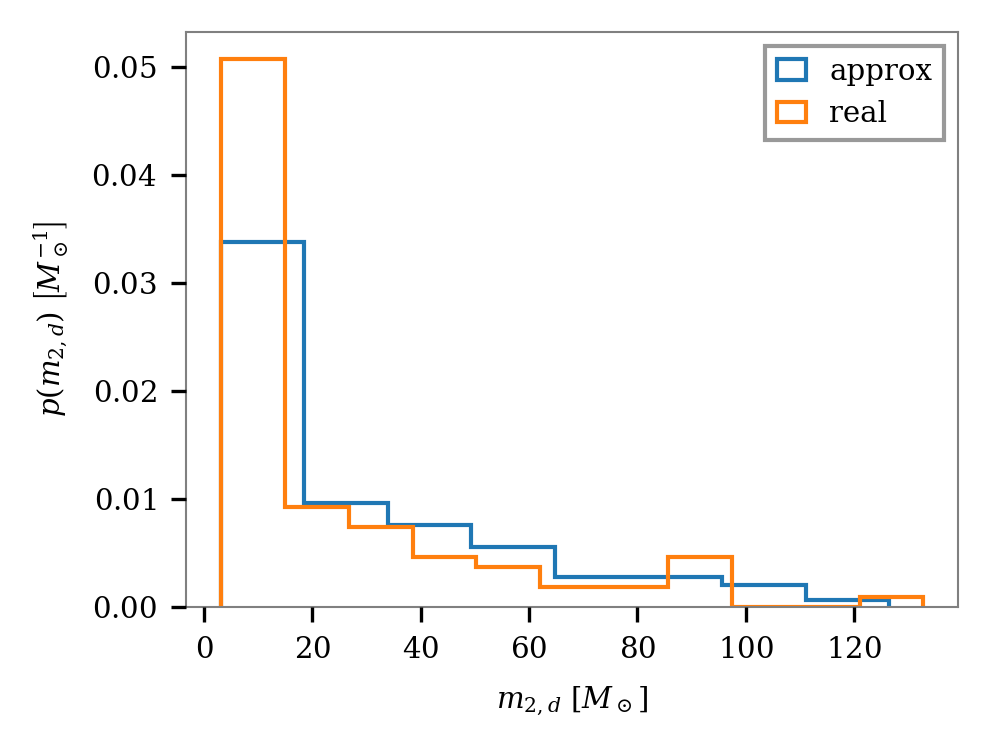}
  \caption{Lighter detector frame mass}
  \label{fig:population m1d}
\end{subfigure}%
\hspace{1em}
\begin{subfigure}[t]{.48\linewidth}
  \centering
\includegraphics[scale=0.95]{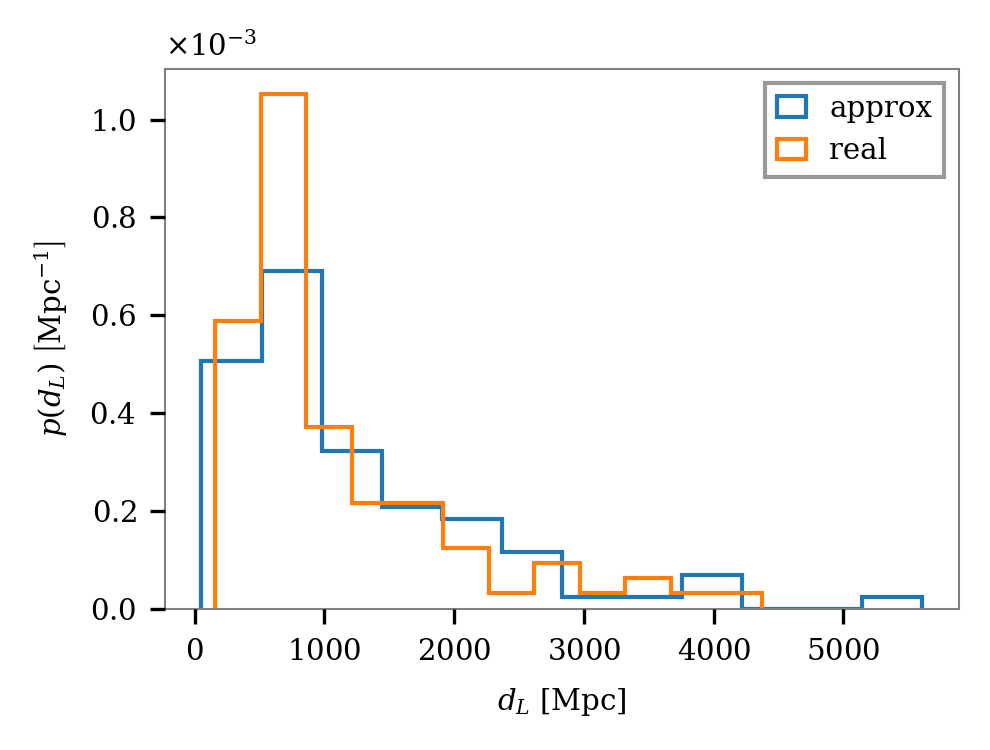}
     \caption{Luminosity distance}
    \label{fig:population m2d}
\end{subfigure}%
   \caption{Distribution obtained for the observed populations after application of the selection criterion. This figure compares the distribution obtained with \texttt{bilby} (orange/\textit{real}) to the one based on the SNR approximation in Eq.\,\eqref{eq: def fit snr} (blue/\textit{approx}). The distribution obtained through the two methods are compatible overall with minor differences in their tails: the distribution based on the SNR approximation has fewer events at lower masses (a) and (b), and goes to further distances (c). All these populations are for O4 sensitivity.}
    \label{fig: approximant test population bilby}
\end{figure}

Let us return to the condition (iii) we have requested above. To compare the detected population with our simplified SNR model and a full calculation of the SNR using the \texttt{IMRPhenomPv2} waveform \cite{Husa:2015iqa,Khan:2015jqa}, we plot Figures \ref{fig:population dl}, \ref{fig:population m1d} \ref{fig:population m2d}. 
We have simulated a population and selected all events that pass the threshold of a network SNR of 12. 
The two populations agree reasonably, but show differences in their tails: 
the SNR approximation allows for detections at higher redshifts. 

\begin{figure}
    \centering
    \begin{subfigure}[t]{.45\linewidth}
  \centering
\includegraphics[width=\linewidth]{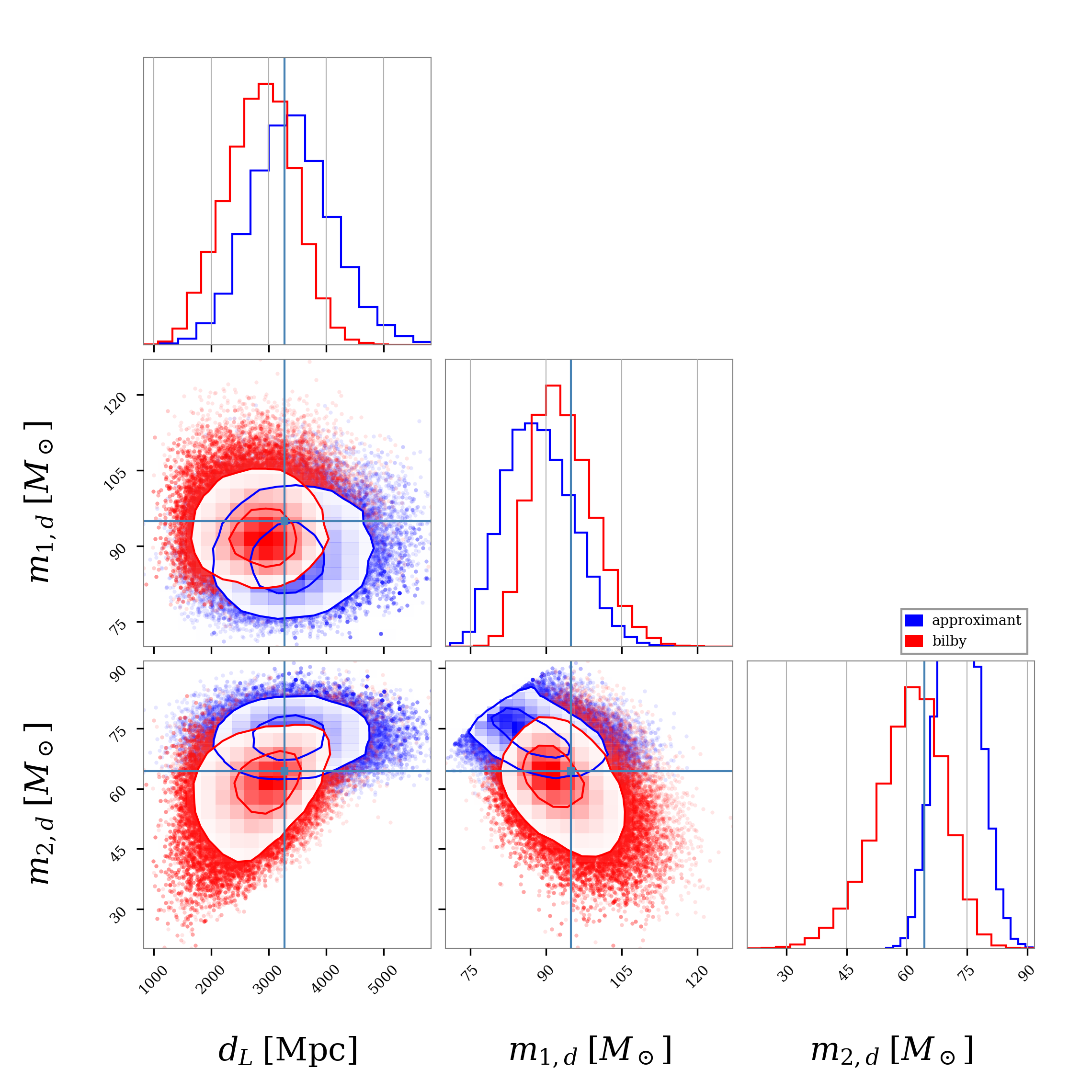}
     \caption{Event 0}
    \label{fig:comparison bilby ev0}
\end{subfigure}%
\hspace{1em}
\begin{subfigure}[t]{.45\linewidth}
  \centering
\includegraphics[width=\linewidth]{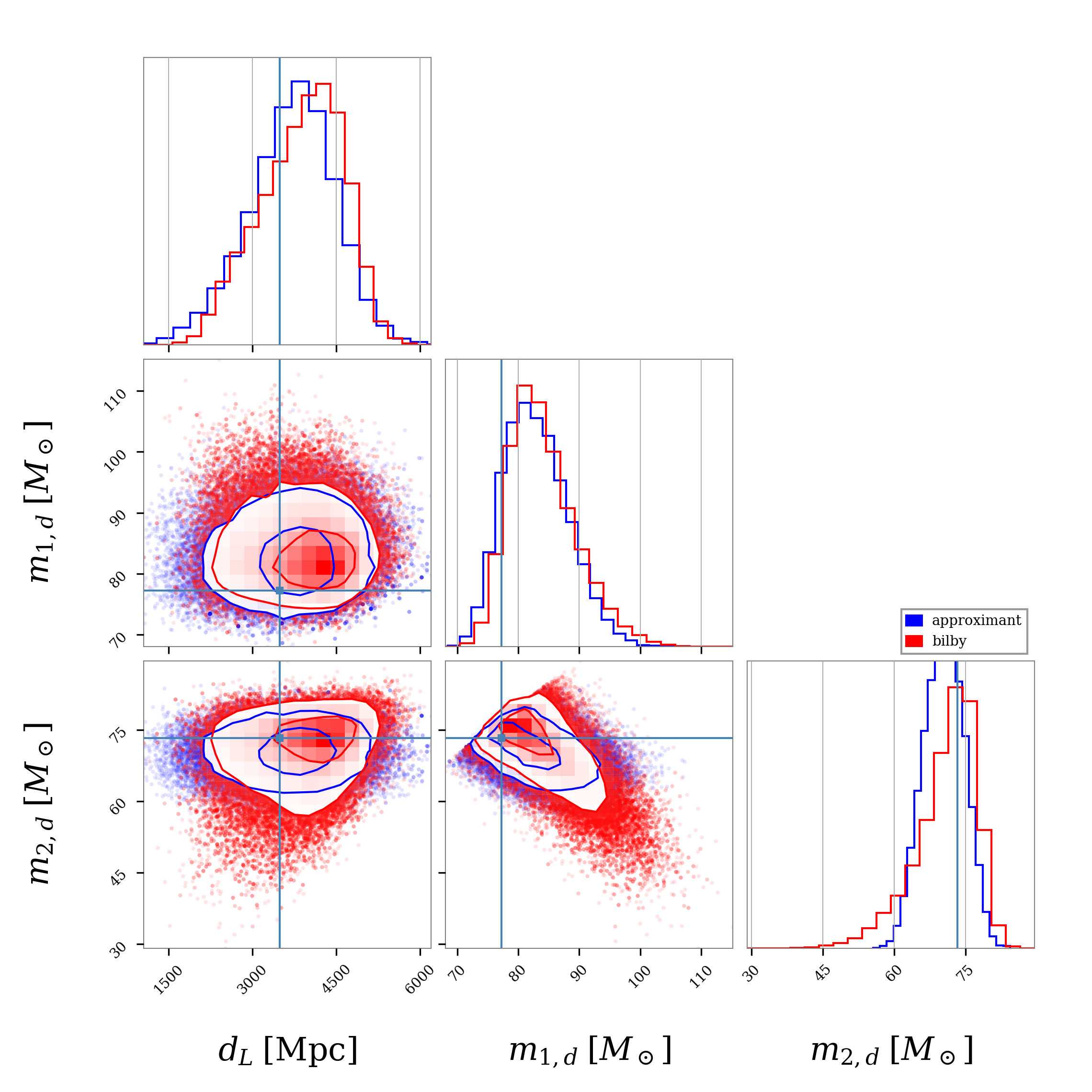}
  \caption{Event 1}
\label{fig:comparison bilby ev1}
\end{subfigure}%
\hspace{1em}
\begin{subfigure}[t]{.45\linewidth}
  \centering
\includegraphics[width=\linewidth]{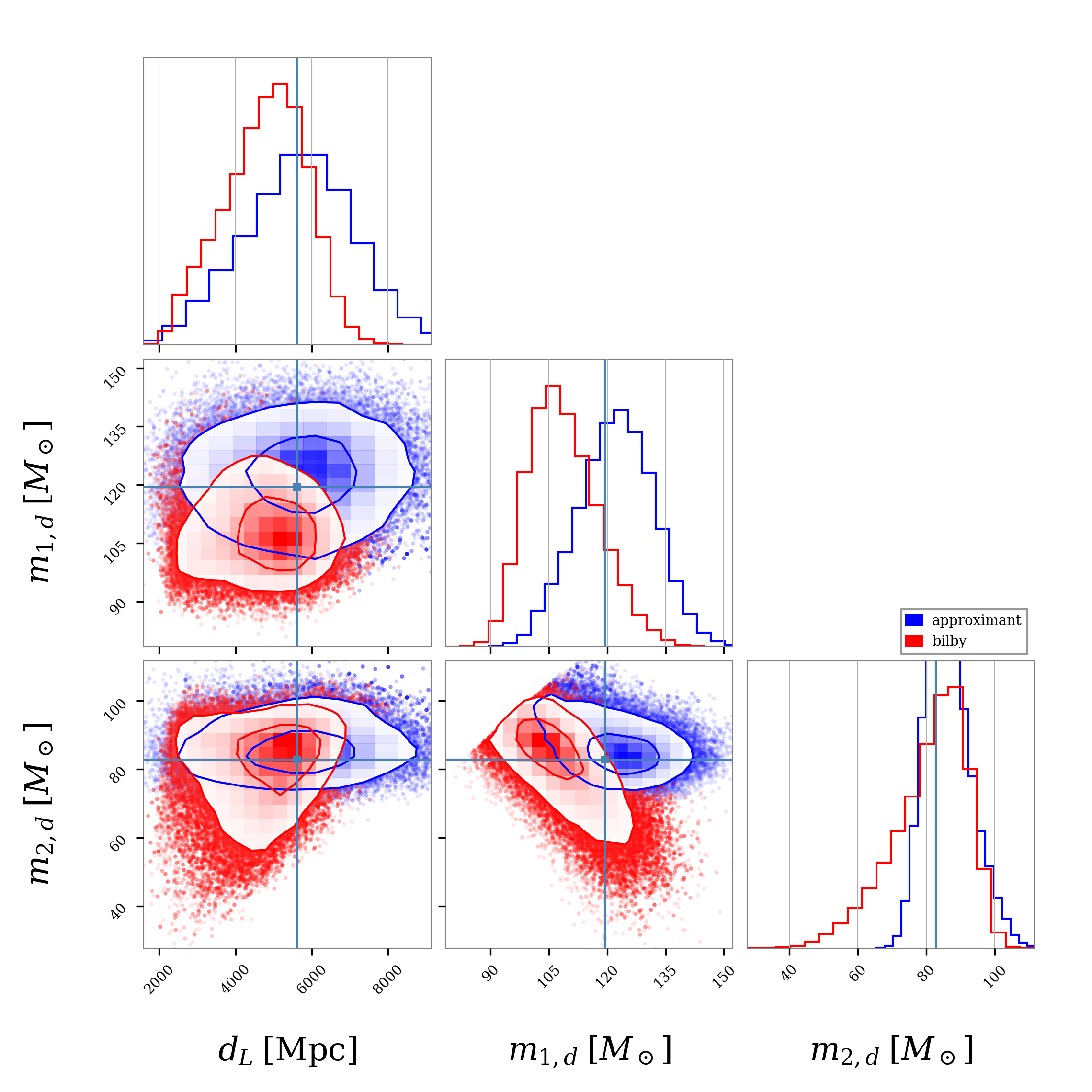}
\caption{Event 2}
\label{fig:comparison bilby ev2}
\end{subfigure}%
\hspace{1em}
\begin{subfigure}[t]{.45\linewidth}
\centering
\includegraphics[width=\linewidth]{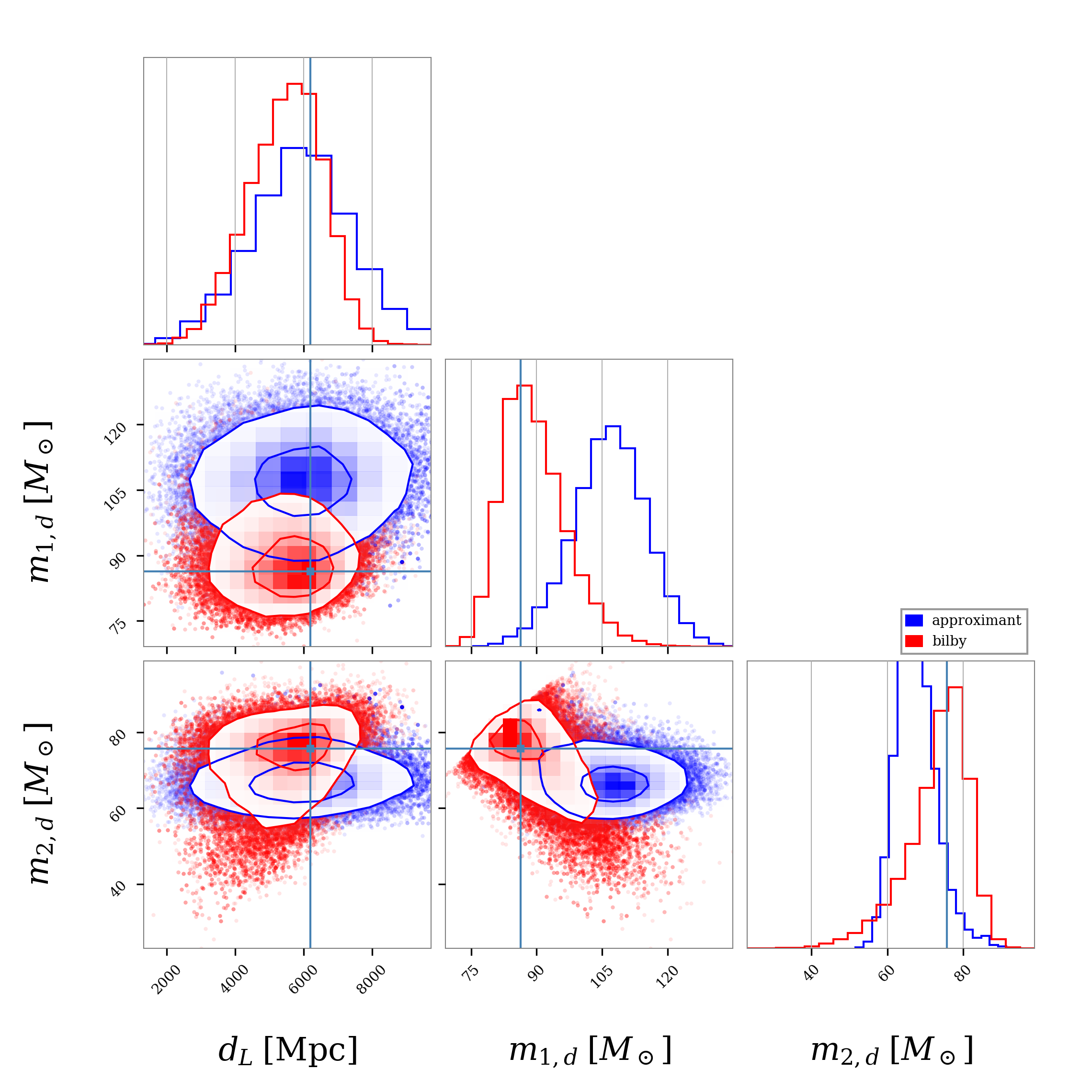}
\caption{Event 3}
\label{fig:comparison bilby ev3}
\end{subfigure}%
\caption{Comparison between a full parameter estimation using \texttt{bilby} (red) and the posterior proxy (blue) for four different GW events.
The posterior distributions obtained for Events 2 in (c) and 3 in (d) slightly disagree in their position (value around which the likelihood is peaked)
but are consistent in terms of scatter.}
\label{fig:comparison bilby}
\end{figure}

Finally, we compare the \texttt{bilby} analysis and the approximant. 
For a number of sources, a full parameter estimation with \texttt{bilby} is performed, using the waveform \texttt{IMRPhenomPv2}, assuming aligned spins and a fixed coalescence time. 
The results for several comparisons can be found in Figures \ref{fig:comparison bilby ev0}, \ref{fig:comparison bilby ev1} ,\ref{fig:comparison bilby ev2} and \ref{fig:comparison bilby ev3}.

\section{Comparison of results obtained with the PL and PLG source frame mass models}
\label{app: results pl vs plg}

\begin{figure}
    \centering
    \includegraphics[width=\textwidth]{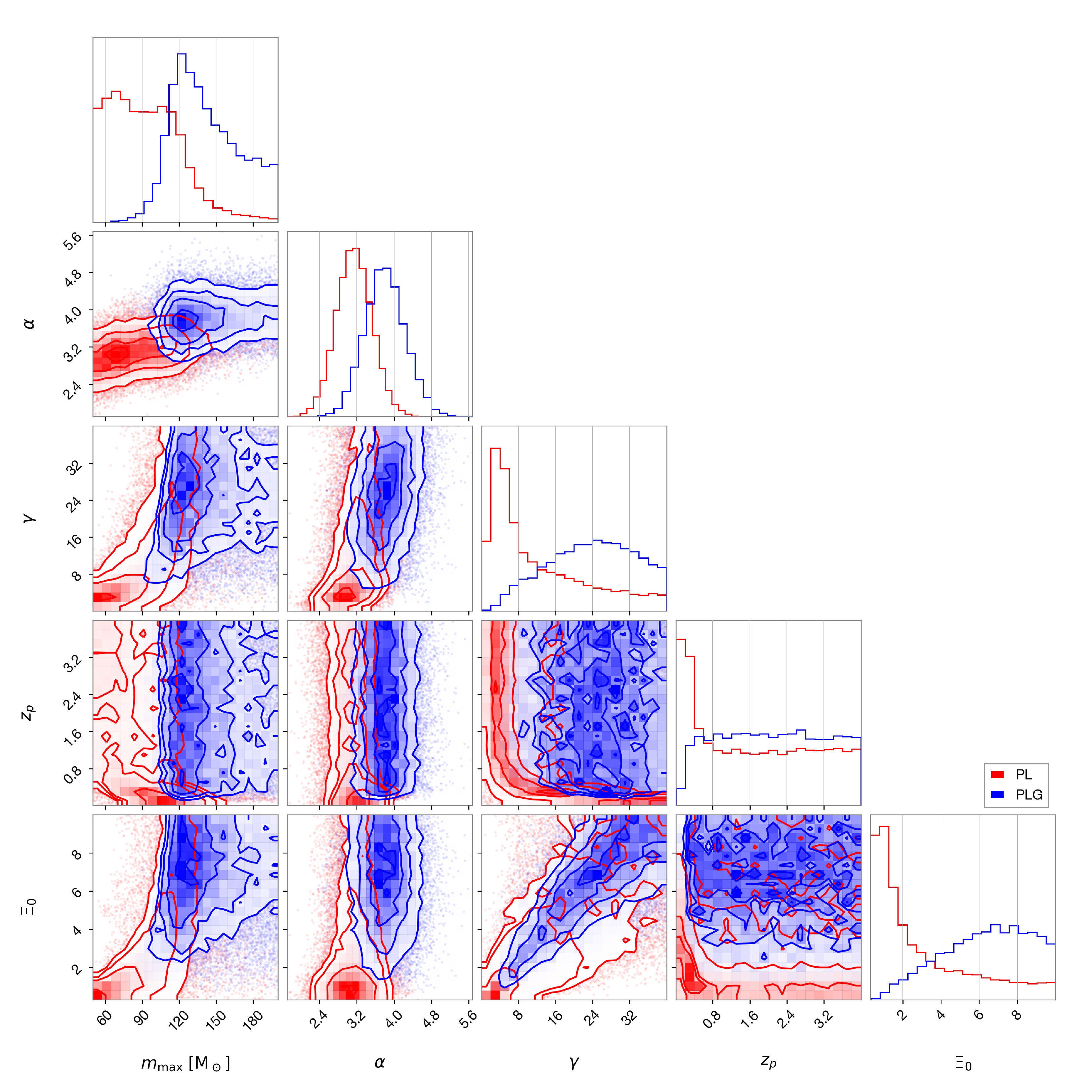}
    \caption{Results obtained with GWTC-3 data using wide priors on $H_0$ and on $\gamma$. The figure compares the posterior distribution based on the PL mass model (red) to the PLG mass model (blue). A SNR cut of 11 and an IFAR of 4 years is applied to the events in the GWTC-3 catalog. Level curves at 1, 2 and 3 sigmas are indicated. }
    \label{fig:o3 pl vs plg}
\end{figure}

\change{Here we elaborate further on the impact on the source frame mass model by comparing the results obtained with the \textsc{truncated} (PL) and \textsc{power law + peak} (PLG) models, see Figure~\ref{fig:o3 pl vs plg}.
As indicated by its lower Bayes factor the PL model is too simple to fit all the features in the observed population. To fit the overdensity in the GWTC-3 event distribution at $\sim$40 solar masses the PL model results in a shallower slope of the primary mass distribution in comparison with the PLG model. PL thus predicts more sources at higher masses than the PLG mass model. The observed numbers of events at high masses being fixed, PL compensates with a lower maximum mass. 
Because of strong correlation between the maximum mass and $\Xi_0$ (see Sec.~\ref{sec:corr_params}), the lower $\mmax$ thus results in a lower $\Xi_0$ for the PL when compared to the PLG.
This appendix thus evidences the strong impact that the choice of mass model can have on the measurement of $\Xi_0$. To obtain a reliable measurement of modified GR parameters it is indispensable to test a range of different mass models and evaluate their goodness-of-fit by comparing their Bayes factor.}

\section{Full results for the O4 \& O5 science runs assuming agnostic cosmological priors}
\label{app: full results agnostic priors}
\begin{figure}
    \centering
    \includegraphics[width=\textwidth]{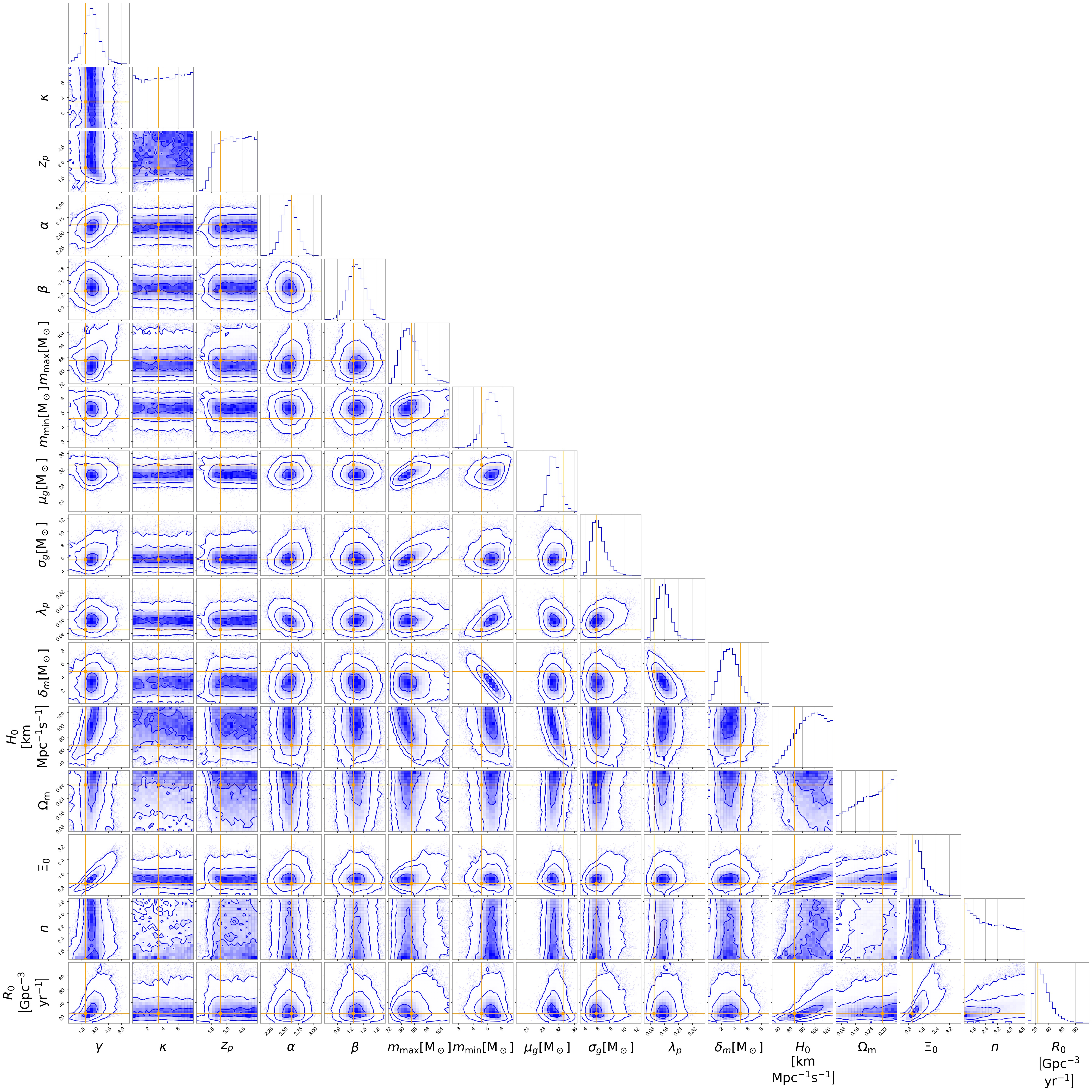}
    \caption{Full results  for O4$+$O5 assuming GR (no friction term) and using wide priors on the cosmology.
    We indicate the level curves of the posterior at 1,2 and 3 sigmas. The true values are shown by orange lines.}
    \label{fig:o4o5 result GR}
\end{figure}

\begin{figure}
    \centering
    \includegraphics[width=\textwidth]{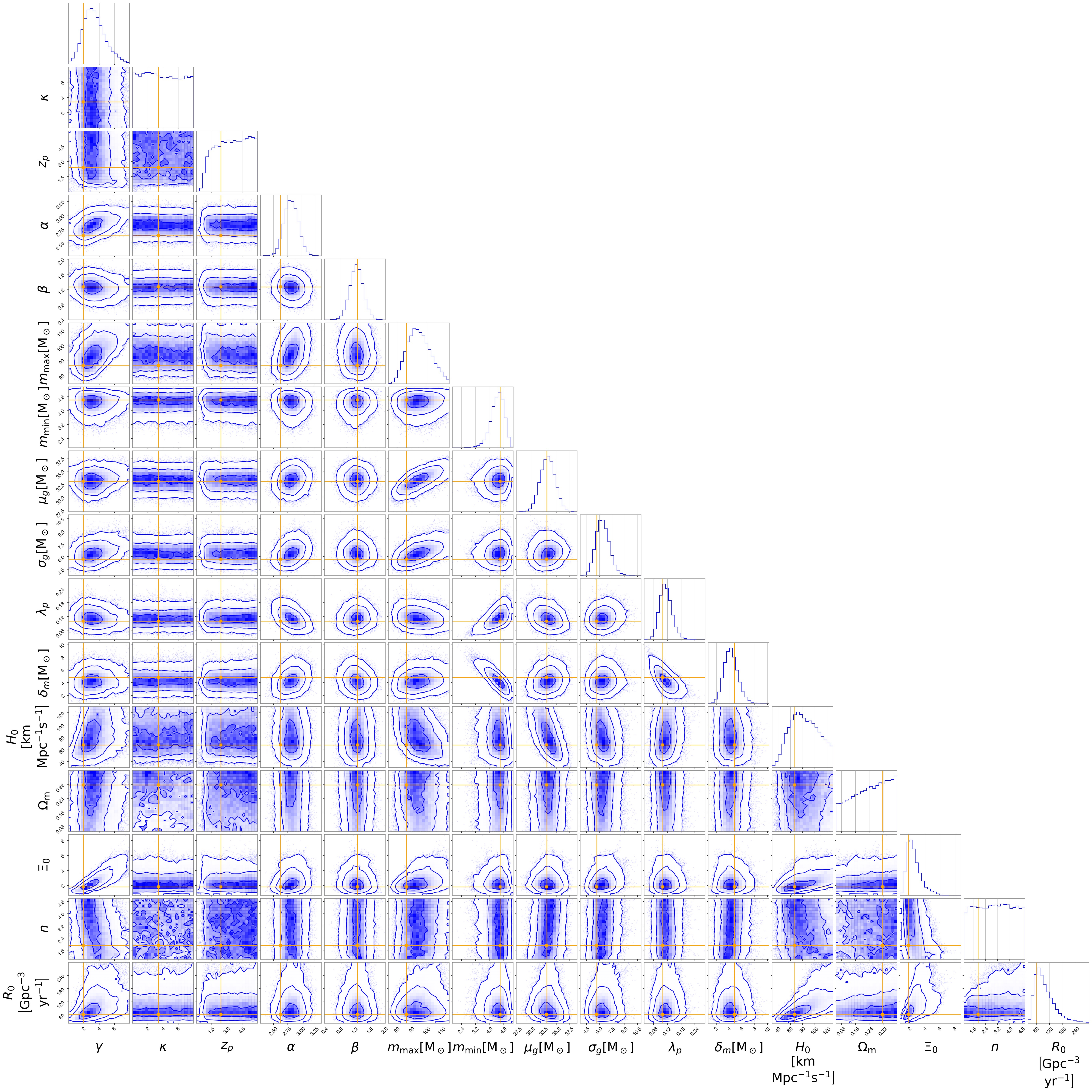}
    \caption{Full results for O4$+$O5 assuming modified gravity and using wide priors on the cosmology.
    We indicate the level curves of the posterior at 1,2 and 3 sigmas. The true values are shown by orange lines.}
    \label{fig:o4o5 result bGR wide}
\end{figure}

In this Appendix, we show some selected corner plots for the \texttt{icarogw} analysis of the posterior for the metaparameters, the analysis which was outlined in Section \ref{sec: bayesian analysis}. 
In Figure \ref{fig:o4o5 result GR} we present the corner plot for the analysis of O4 and O5 produced with a simulated data set based on a flat $\Lambda$CDM universe with no friction term (GR case) and using agnostic priors on the cosmology. 
Figure \ref{fig:o4o5 result bGR wide} is obtained with synthetic data that models a universe with a modified GW propagation. 
This figure shows the analysis for O4 and O5, with broad priors on the cosmological parameters.
The results of Sections \ref{sec:52} and \ref{sec:53} are based on the posteriors presented in these two figures. 

\section{Analysis for the O4 \& O5 science runs with the GR model}
\label{sec:result_O4-O5_GR}

Here we study the expected precision with which we expect to 
recover cosmological parameters when both the simulated dataset and the analysis
model is based on GR (i.e., the modified gravity parameters are fixed to their
default GR values). 

\begin{figure}
    \centering
\includegraphics[width=0.9\textwidth]{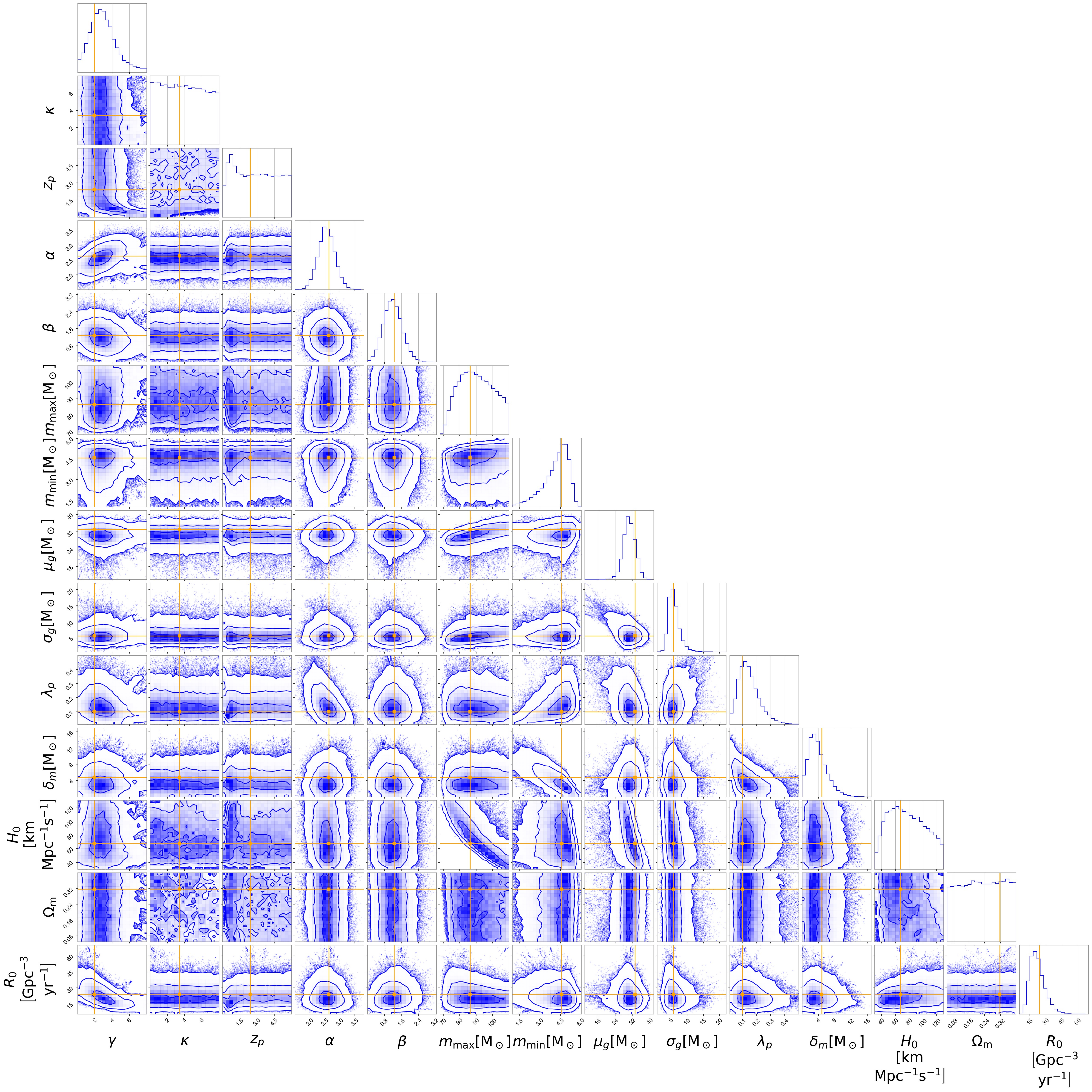}
    \caption{Results obtained for a simulated O4 science run (87 detected events) with the GR model. (Modified gravity parameters are fixed in this analysis to their default GR value).}
    \label{fig:o4 icarogw fix GR wide}
\end{figure}

\begin{figure}
    \centering
    \includegraphics[width=0.9\textwidth]{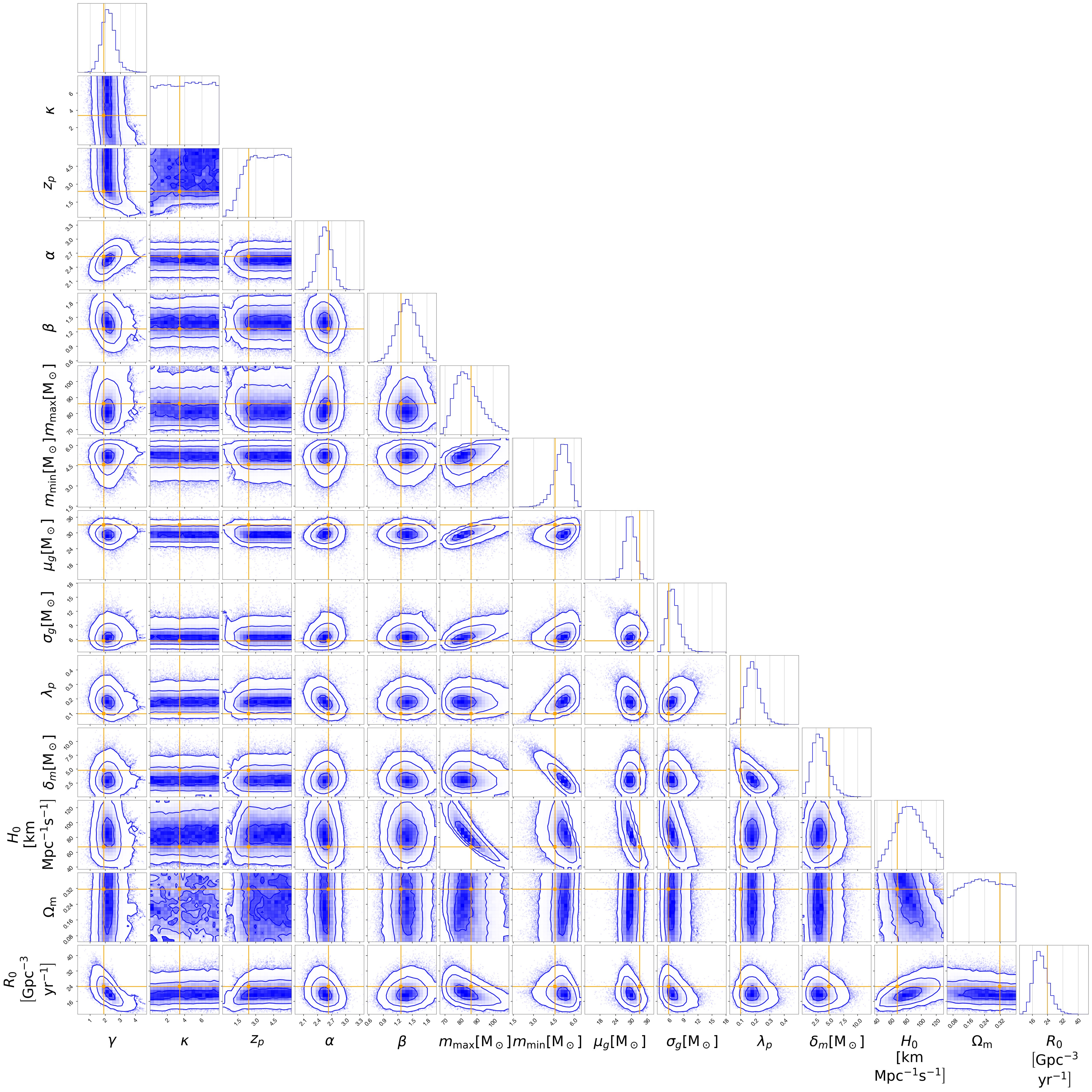}
    \caption{Results obtained for a simulated O5 science run (423 detected events) with the GR model. (Modified gravity parameters are fixed in this analysis to their default GR value).}
    \label{fig:o5 icarogw fix GR wide}
\end{figure}

Fig.~\ref{fig:o4 icarogw fix GR wide} and \ref{fig:o5 icarogw fix GR wide} show the results obtained with 
simulated O4 and O5 science runs, leading to 87 and 423 detected events respectively. 
In both cases, the matter content of the universe $\Omega_{\mathrm{m}}$ is essentially unconstrained. We find that
$H_0 = 76^{+ 32}_{-26}\hu$ for O4 and $H_0=85^{+21}_{-20}\hu$ (i.e., 24 \% {precision}) for O5.
This is only a marginal improvement to the Hubble constant measurement (26 \% {precision}) when $\Xi_0$ and $n$ were left to vary. 
When comparing our results obtained for O5 with those in \cite{Farr:2019twy} ($12\%$ after one year of observation\footnote{This 
uncertainty was obtained by adapting and running the published code used by \cite{Farr:2019twy}.}), 
we find larger uncertainties on the Hubble constant by a factor of $\sim 2$. However, it is difficult to make clear conclusions as those comparisons are not strictly "apple-to-apple".
The analysis presented here assumes a SNR cutoff of 12, while \cite{Farr:2019twy} assumes a cutoff of 8, resulting in a factor of 2.5 difference in the number of observed events. 
With 5 years of observation time, the Hubble constant measurement in \cite{Farr:2019twy} only marginally improves in precision to $11\%$, an error decay slower that the standard asymptotic law $\propto 1/\sqrt{N_{\rm obs}}$. 
For a pivotal redshift of $z\approx0.7$ we find an uncertainty of $H(z)$ of $23\%$ which is much larger than the value of $6\%$ in \cite{Farr:2019twy} after one year of observation.
As \cite{Mancarella:2021ecn} points out, \cite{Farr:2019twy} uses a redshift evolution model that only depends on one parameter, which can decrease the $H_0$ error bars by a factor of 2. 
Compared to \cite{Mancarella:2021ecn}, we find a difference of $20\%$ in the predicted error bars for $H_0$. 
The population in \cite{Mancarella:2021ecn} is assumed to be a \textsc{broken power law}, whereas we assume a \textsc{power law + peak} mass population. 

\section{Posterior distribution of \texorpdfstring{$\Xi_0$}{} for different realizations of the event catalog}
\label{sec:xi0_catalog-realization}

\begin{figure}
    \centering
    \includegraphics{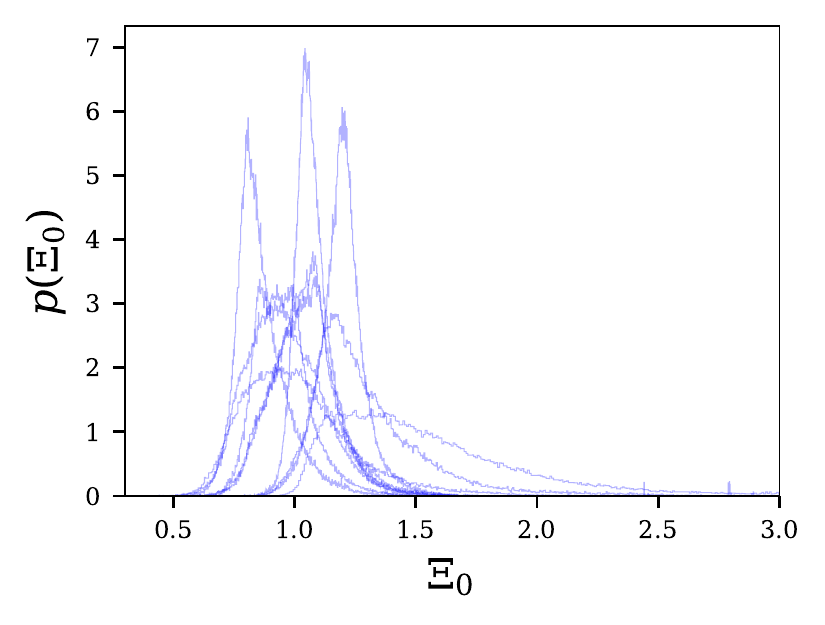}
    \caption{$\Xi_0$ posterior obtained for 10 different realizations of the BBH population with the same metaparameters. 
    To make this check feasible computationally, the uncertainties on the intrinsic GW parameters (masses and GW luminosity distance) were neglected.}
    \label{fig:identical pop parameters delta }
\end{figure}

In order to understand the fluctuations for \textit{fixed population parameters}, we perform several population analyses, where the underlying population is fixed to the parameters of Section \ref{sec: forecast o4 o5}. We assume a GR universe and a HLV network with a sensitivity and observation times as given in Section \ref{sec: forecast o4 o5}.   
Fig.\,\ref{fig:identical pop parameters delta } presents the posteriors on $\Xi_0$ for 10 identical runs, where we assume 510 detected events. 
For each new run, we generate a new catalog of events. The $\Xi_0$ posterior displays a scatter depending on the realizations of the population. 

\acknowledgments

The authors are grateful for computational resources provided by the LIGO Laboratory and supported by the National Science Foundation Grants PHY-0757058 and PHY-0823459.
This research has made use of data or software obtained from the Gravitational Wave Open Science Center (gw-openscience.org), a service of LIGO Laboratory, the LIGO Scientific Collaboration, the Virgo Collaboration, and KAGRA. LIGO Laboratory and Advanced LIGO are funded by the United States National Science Foundation (NSF) as well as the Science and Technology Facilities Council (STFC) of the United Kingdom, the Max-Planck-Society (MPS), and the State of Niedersachsen/Germany for support of the construction of Advanced LIGO and construction and operation of the GEO600 detector. Additional support for Advanced LIGO was provided by the Australian Research Council. Virgo is funded, through the European Gravitational Observatory (EGO), by the French Centre National de Recherche Scientifique (CNRS), the Italian Istituto Nazionale di Fisica Nucleare (INFN) and the Dutch Nikhef, with contributions by institutions from Belgium, Germany, Greece, Hungary, Ireland, Japan, Monaco, Poland, Portugal, Spain. The construction and operation of KAGRA are funded by Ministry of Education, Culture, Sports, Science and Technology (MEXT), and Japan Society for the Promotion of Science (JSPS), National Research Foundation (NRF) and Ministry of Science and ICT (MSIT) in Korea, Academia Sinica (AS) and the Ministry of Science and Technology (MoST) in Taiwan.
Numerical computations were performed on the DANTE platform, APC, France. 
We would like to thank Marco Crisostomi for helpful discussion on the computation of modified luminosity distances in DHOST models.
The Fondation CFM pour la Recherche in Paris has generously supported KL during his doctorate.
SM is supported by the ANR COSMERGE project, grant ANR-20-CE31-001 of the French Agence Nationale de la Recherche. SM acknowledges the support of the LabEx UnivEarthS (ANR-10-LABX-0023 and ANR-18-IDEX0001), of the European Gravitational Observatory and of the Paris Center for Cosmological Physics.

\bibliographystyle{JHEP}
\bibliography{references}

\end{document}